\documentclass{article}

\usepackage{arxiv}

\usepackage[utf8]{inputenc} % allow utf-8 input
\usepackage[T1]{fontenc}    % use 8-bit T1 fonts
\usepackage{url}            % simple URL typesetting
\usepackage{booktabs}       % professional-quality tables
\usepackage{amsfonts}       % blackboard math symbols
\usepackage{nicefrac}       % compact symbols for 1/2, etc.
\usepackage{microtype}      % microtypography
\usepackage{graphicx}
\usepackage{natbib}
\usepackage{doi}
\usepackage[all]{xy}
\usepackage{setspace}
\usepackage{xcolor}
\usepackage{tikz}
\usetikzlibrary{cd}
\usepackage{amsmath}
\usepackage{bm}
\usetikzlibrary{shapes, arrows.meta, positioning, shadows}
\usepackage{mathtools}

\newcommand{\UU}{\mathbf{U}}
\newcommand{\gmetric}{\mathbf{g}}

\newcommand{\uS}{{^\uparrow S}}
\newcommand{\dS}{{^\downarrow S}}
\newcommand{\dF}{{^\downarrow F}}
\newcommand{\dSt}{{^\downarrow \tilde{S}}}
\newcommand{\uE}{{^\uparrow E}}
\newcommand{\dE}{{^\downarrow E}}
\newcommand{\qq}{\mathbf{q}}
\newcommand{\pp}{\mathbf{p}}
\newcommand{\QQ}{\mathbf{Q}}
\newcommand{\PP}{\mathbf{P}}
\newcommand{\xx}{\mathbf{x}}
\newcommand{\XX}{\mathbf{X}}
\newcommand{\dX}{{{^\downarrow}X}}
\newcommand{\dXX}{{{^\downarrow}\XX}}
\newcommand{\yy}{\mathbf{y}}
\newcommand{\YY}{\mathbf{Y}}
\newcommand{\dL}{{{^\downarrow}L}}
\newcommand{\WW}{\mathbf{W}}
\newcommand{\PPi}{\boldsymbol{\Pi}}
\newcommand{\rr}{\mathbf{r}}

\newcommand{\xxi}{\boldsymbol{\xi}}

\title{How the arrow of time emerges from Hamiltonian systems by our incomplete knowledge}

\author{Kateřina Mladá\\
Mathematical Institute, Faculty of Mathematics and Physics, Charles University,\\ 
 Sokolovsk\'{a} 83, 18675 Prague, Czech Republic
\\
\And Michal Pavelka\\
Mathematical Institute, Faculty of Mathematics and Physics, Charles University,\\ 
 Sokolovsk\'{a} 83, 18675 Prague, Czech Republic\\
Corresponding author: pavelka@karlin.mff.cuni.cz\\
\And Václav Klika\\
	Dept. of Mathematics, FNSPE, 
Czech Technical University in Prague,\\
Trojanova 13, 120 00 Prague, Czech Republic
}

\newcommand{\shorttitle}{Arrow of time from incomplete knowledge: a path-integral approach}

\usepackage{amsmath,amssymb,amsfonts,latexsym,float,graphics,epsfig}

\begin{document}
\maketitle

\begin{abstract}
How does the arrow of time (dissipative, irreversible behavior) emerge from time-reversible Hamiltonian mechanics? Two ingredients are needed: the underlying system must be ergodic or phase-mixing, and our knowledge of the system must be incomplete. When the detailed dynamics explores its phase space and stays close to a submanifold parametrized by a reduced set of state variables, the lack-of-fit reduction method reveals that the effective equations for those reduced variables are necessarily irreversible. To make this precise, we present a path-integral formulation of the lack-of-fit reduction in non-equilibrium thermodynamics, which shows how the GENERIC framework (reversible Hamiltonian part plus irreversible gradient flow) emerges from purely Hamiltonian mechanics without any fitting parameters. The formulation is based on the Onsager-Machlup variational principle, and it yields reduced dynamical equations by minimizing the information discrepancy between the detailed and reduced evolutions. Subsequently, the reduction method is illustrated on the Kac--Zwanzig model, confirming that dissipation emerges from ignoring degrees of freedom, and on diffusion, where diffusion equation for the hydrodynamic mass density emerges from Vlasov equation. We also show how to generalize the Fisher information matrix and Kullback--Leibler divergence to arbitrary concave entropies via the principle of maximum entropy, including non-Boltzmann-Gibbs cases such as the Tsallis--Havrda--Charv\'{a}t entropy.
\end{abstract}

\tableofcontents

\doublespace

\subsection*{Notation}
\begin{table}[h!]
\centering
\begin{tabular}{ll}
\toprule
\textbf{Symbol} & \textbf{Meaning} \\
\midrule
$\uparrow$, $\downarrow$ & Superscripts denoting the detailed (upper) and reduced (lower) level of description \\
$\xx$, $\yy$ & Detailed and reduced state variables \\
$\yy^*$ & Conjugate (dual) reduced variables, obtained by differentiating entropy \\
$\pi:\xx\mapsto\yy$ & Projection from detailed to reduced variables \\
$\mu:\yy^*\mapsto\xx(\yy^*)$ & MaxEnt mapping from conjugate reduced variables to detailed variables \\
$\uS$, $\dS$ & Entropy at the detailed and reduced levels \\
$\dSt^*$ & Reducing potential (Eq.~\eqref{eq.dSt*}) \\
$\dS^*$ & Reduced conjugate entropy (Legendre transform of $\dS$) \\
$D^{M*}_{KL}$ & MaxEnt Kullback--Leibler discrepancy (Eq.~\eqref{eqn:DKLM}) \\
$\gmetric$, $g^{ab}$ & Fisher information matrix / metric on the reduced state space (Eq.~\eqref{eq.gab}) \\
$\XX$, $\YY$ & Detailed and reduced evolution vector fields \\
$\dXX$ & Projection of the detailed vector field onto the reduced space \\
$\mathcal{L}$ & Lack-of-fit Lagrangian (Eq.~\eqref{eqn:Lagrangian_GENERIC}) \\
$\Sigma_e$ & Extremal action / emergent dissipation potential \\
$\uE$, $\dE$ & Energy at the detailed and reduced levels \\
${^\uparrow}L^{ij}$, ${^\downarrow}L^{ab}$ & Poisson bivector at the detailed and reduced levels \\
${^\uparrow}\Xi$ & Dissipation potential at the detailed level \\
\bottomrule
\end{tabular}
\end{table}

\section{Introduction}
How the arrow of time (irreversible, dissipative macroscopic evolution) emerges from the time-reversible equations of microscopic mechanics \cite{boltzmann1872,boltzmann1896,gibbs1902}. This problem, known classically as the conflict between the time-reversibility of microscopic dynamics and the time-irreversibility of macroscopic behavior \cite{loschmidt1876,poincare1890}, remains central to our understanding of thermodynamics and statistical mechanics. In this paper, we argue that two ingredients are needed to resolve this apparent paradox. First, the underlying system must be ergodic or phase-mixing \cite{sinai1963,arnold-avez1968}, so that the detailed trajectory explores its phase space in a statistically well-defined way. Second, our knowledge must be incomplete: only a reduced set of state variables is tracked and the remaining degrees of freedom are ignored \cite{kubo,evans-searles,netocny2002}. Given both conditions, the lack-of-fit reduction method reveals that the effective equations governing the reduced variables are irreversible --- the arrow of time is not put in by hand, but derived.

The purpose of non-equilibrium thermodynamics is to provide a framework for describing the evolution of physical systems. For that purpose, it is necessary to have reduction methods, as any system can be described at various levels of description with different amounts of detail. In this paper, we compute the information discrepancy between a detailed and reduced description of a physical system, and by minimization of that discrepancy we obtain a set of reduced evolution equations.

The detailed evolution is assumed to be in the form of the General Equation for Non-Equilibrium Reversible-Irreversible Coupling (GENERIC), where the evolution is the sum of a reversible (Hamiltonian) part and irreversible (gradient) part \cite{go,og,hco,pkg}. The reversible part is generated by a Poisson bracket and the Hamiltonian, while the irreversible part is generated by a dissipation potential and entropy. The reversible part is assumed to be purely mechanical, while the irreversible part is dissipative. In particular, the detailed evolution can also be purely Hamiltonian, while the reduced evolution will always be dissipative. 

A general reduction method in non-equilibrium thermodynamics should satisfy the following criteria: (i) it should be compatible with the principle of maximum entropy, (ii) it should not require any fitting parameters, and (iii) it should be able to reduce a purely Hamiltonian system to a dissipative one. The lack-of-fit reduction \cite{JSP2020,lof2024} satisfies all these criteria. The method is based on the Onsager-Machlup variational principle \cite{om}, which gives geodesics of the action (lack of fit between the reduced and detailed evolutions).

Another criterion for a good reduction method is the closeness of the reduced and detailed evolutions \cite{zwanzig}. Here, we hypothesize that such closeness should automatically follow, assuming that the detailed dynamics stays close to the MaxEnt submanifold \cite{gk}, that is the manifold of states obtained by the principle of maximum entropy (MaxEnt). The lack-of-fit reduction then minimizes the discrepancy between the detailed vector field and the reduced vector field on that submanifold.

The lack-of-fit reduction originates from works of Bruce Turkington and co-workers \cite{turkington},
 \cite{Turkington-turbulence}, \cite{Turkington-point-vortex}, \cite{Turkington-fluctuations}, where a variational principle was proposed that measures the discrepancy between the detailed and reduced dynamics. The method was further developed by Richard Kleeman \cite{kleeman}, who added a path-integral formulation and a stochastic extension. Although in the original formulation, the method required fitting parameters, later versions were parameter-free, with applications in turbulence modelling. In the current manuscript, we provide an information-geometric formulation the lack-of-fit reduction \cite{JSP2020}, using the path-integral formulation of \cite{kleeman}. We also show how the precision of the reduced evolution can be measured by the Kullback–Leibler discrepancy between the detailed and reduced evolutions for general entropies, and we show how the diffusion constant of a gas can be derived from the lack-of-fit reduction of the Vlasov equation, which is a purely Hamiltonian system.

There are several other reduction methods in non-equilibrium thermodynamics. The Mori-Zwanzig projection operator method \cite{Mori65,zwanzig,hco} rewrites a second-order differential equation into an integro-differential first-order form with a memory kernel. The kernel is then approximated to obtain the reduced evolution \cite{grabert}. Although this method often leads to a good approximation of the reduced evolution, characterized by a much larger time-scale than the detailed evolution, it often requires measurement of the fluctuations or a fitting parameter \cite{grabert}, and it does not always lead to a dissipative term in the evolution. 

The Ehrenfest reduction \cite{gk-ehrenfest,gk-ehrenfest2,ehre} is another reduction method based on an approximate dissipative solution of the detailed system. However, it requires a fitting parameter that expresses the relaxation time of the detailed dynamics towards the manifold of reduced variables \cite{gk}. More recently, an interesting approach was proposed for coarse-graining of a dissipative Langevin equation \cite{reina-nexus}, which shows remarkable robustness, and another approach was proposed based on finding a dissipative semigroup with an a priori known dissipation parameter \cite{mielke2025}.

To formulate the information discrepancy for arbitrary concave entropy functionals, we generalize the Kullback–Leibler (KL) divergence and the Fisher information matrix to arbitrary entropies, using the principle of maximum entropy \cite{kl,fisher,jaynes}. Finally, we obtain a simpler version of the lack-of-fit reduction than before \cite{JSP2020}, and we test it on the Kac–Zwanzig model, including the stochastic version of the reduced dynamics. 

In Section \ref{sec.generic}, we recall the GENERIC framework, and Section \ref{sec.KL} is devoted to the generalized KL divergence (Kullback-Leibler discrepancy) and the Fisher information matrix. In Section \ref{sec.lof}, we present the lack-of-fit reduction, and in Section \ref{sec.path_integral}, we give its path-integral formulation. In Section \ref{sec.KZ}, we apply the method to the Kac–Zwanzig model, and we show how different choices of the reduced state variables agree with the detailed numerical simulations, based on their KL divergence with respect to the distribution of the detailed state variables. Finally, in Section \ref{sec.diffusion}, we illustrate the lack-of-fit reduction on the example of diffusion emerging from the Vlasov equation when interactions between particles are present.

\section{GENERIC framework}\label{sec.generic}
Within the GENERIC framework \cite{go,og,hco,pkg}, evolution of any set of state variables $\xx$ is the sum of two contributions, one of which is \emph{Hamiltonian} (reversible) and the other is \emph{dissipative} (irreversible),
\begin{equation}
    \dot{\xx} = {^{\uparrow}}\{\xx,\uE\} + \frac{\delta ^{\uparrow}  \Xi}{\delta \xx^*}\Big|_{\xx^* = \frac{\delta \uS}{\delta \xx}}.
\end{equation}
Throughout this manuscript, we use $\uparrow$ to denote quantities related to the detailed state variables $\xx$ (higher level of detail, higher level of description), while $\downarrow$ denotes quantities related to the reduced state variables $\yy$ (lower level of description). The Hamiltonian (reversible) evolution is generated by a Poisson bracket ${^{\uparrow}}\{,\}$ and the Hamiltonian $\uE$ (the energy). The dissipative part is generated by a dissipation potential ${^{\uparrow}}\Xi$ and entropy $\uS$ (the conjugate variables $\xx^*$ are obtained from the entropy $\uS(\xx)$ by the functional derivative). The Poisson bracket is bilinear, skew-symmetric, and it satisfies the Leibniz rule and the Jacobi identity. Skew-symmetry ensures that the energy is conserved by the Hamiltonian evolution. The Leibniz rule makes the bracket work like a derivative, so the evolution equations are invariant with respect to adding constants to the energy. Before explaining the Jacobi identity, we shall first recall the concept of Poisson bivector.

The GENERIC evolution can be also seen as being generated by a vector field $\XX$, components of which have the form
\begin{equation}\label{eq.X}
X^i = {^\uparrow} L^{ij}  \frac{\partial \uE}{\partial x^j} + \frac{\partial ^{\uparrow}  \Xi}{\partial x^*_i} \Bigg|_{\mathbf{x}^* = \frac{\partial \uS}{\partial {\mathbf{x}}}}
\end{equation}
where
\begin{equation}
    ^\uparrow L^{ij} = {^\uparrow}\{x^i,x^j\}
\end{equation}
are the components of a \emph{Poisson bivector} (actually a twice contravariant tensor field). The Jacobi identity can be then interpreted as that the Lie derivative of the bivector with respect to the Hamiltonian part of the vector field is zero \cite{fecko,lageul}.

The dissipation potential ${^{\uparrow}}\Xi$ depends on the conjugate variables such that it has a minimum at $\xx^*=0$, and it is convex near that point. It should not change the overall energy of the system, for which a \emph{degeneracy condition} is needed \cite{kraaj}. Moreover, it can also depend directly on the state variables $\xx$, which is needed for instance in the case of chemical reactions \cite{grmela-loma,mielke-potential}. The dissipation potential generates a \emph{gradient flow}, which can also be seen as the most probable path of a stochastic system obeying a large-deviation principle \cite{mielke-potential,kraaj,hco-jnet2020-I,hco-jnet2020-II}. 

When we have a reduced set of state variables $\yy$ obtained by a projection $\pi:\xx\mapsto\yy$, the Poisson bivector can be projected onto the reduced state space as
\begin{equation}
    ^\downarrow L^{ab} = {^\uparrow}\{y^a,y^b\} = \frac{\partial y^a}{\partial x^i} \frac{\partial y^b}{\partial x^j} \left( ^\uparrow L^{ij} \right).
\end{equation}
If the resulting $^\downarrow L^{ab}$ is independent of $\xx$, the corresponding bracket satisfies the Jacobi identity. In other cases, another mapping $\xx(\yy)$ might be necessary to remove that dependence, and it is often the maximum entropy mapping, discussed in the following section. 

The reduced evolution can also be expressed in a dual form, where conjugate variables $\yy^*$ are obtained as Legendre transformations of $\yy$, and the Poisson bivector then transforms as a twice contravariant tensor field,
\begin{equation}
    ^\downarrow L^*_{ab} = {^\uparrow}\{y^*_a,y^*_b\}|_{\xx(\yy^*)} = \frac{\partial y^*_a}{\partial x^i} \frac{\partial y^*_b}{\partial x^j} \left( ^\uparrow L^{ij} \right)|_{\xx(\yy^*)}.
\end{equation}
These forms of the Poisson bivector will help us later when deriving the reduced evolution equations by means of the lack-of-fit reduction.

A similar structure as GENERIC can be found in \cite{dv,be}, where entropy density is among the state variables. When the dissipation potential is quadratic, its Hessian plays the role of a metric, and the evolution equations then obtain the \emph{metriplectic} form \cite{mor,grcontmath}, where the dissipative part can also be expressed using a dissipative bracket \cite{hco}, which in some generalizations even turns out not fully symmetric \cite{hco}. GENERIC with a quadratic dissipation potential can also be seen as essentially equivalent with the \emph{Steepest Entropy  Ascent} (SEA) method \cite{sea,sea-generic}.

\section{Kullback--Leibler discrepancy: A generalization of Kullback–Leibler divergence}\label{sec.KL}
Before formulating the lack-of-fit reduction, we have to recall the concept of Kullback–Leibler divergence, and we have to generalize it so that we can work with an arbitrary concave entropy $\uS$. The resulting Kullback--Leibler discrepancy will then measure the distance between elements $\xx$ and $\yy$ of two different manifolds related by the principle of maximum entropy.

\subsection{Classical Kullback–Leibler divergence}
The Kullback–Leibler (KL) divergence is a measure of the difference between two probability distributions \cite{kl}. It is defined as
\begin{equation}\label{eq.KL}
D_{KL}(\pp || \qq) = \sum_i p_i \ln{\frac{p_i}{q_i}} 
\end{equation}
where $\pp$ and $\qq$ are two probability distributions. The KL divergence is non-negative, as can be proved by noting that $\ln(x) \leq x-1$, and is equal to zero if and only if the two distributions are identical.

The form of KL divergence is strongly tied with the Shannon entropy 
\begin{equation}\label{eq.shannon}
    \uS(\pp) = -k_B \sum_i p_i \ln(\alpha p_i),
\end{equation}
where $\alpha$ is a constant (typically equal to the unity for discrete distribution and to a power of the Planck constant for distribution on a phase space) \cite{boltzmann,shannon,pkg}. 

\subsection{Principle of maximum entropy}
Let us first recall the geometric formulation of MaxEnt \cite{redext}. Assume that a detailed set of state variables $\xx$ is equipped with entropy $\uS(\xx)$, which is a function of the state variables. A reduced set of state variables $\yy$ is obtained by a projection $\pi:\xx\mapsto\yy$. The \emph{MaxEnt estimate} of $\xx$ based on the knowledge of $\yy$ is obtained in two steps. 

First, the MaxEnt estimate is calculated as a function of the reduced conjugate variables $\yy^*$ by
\begin{equation}
    \frac{\partial}{\partial \xx}\left( -\uS(\xx) + y^*_a \pi^a(\xx)\right) = 0,
    \label{eqn:MaxEnt}
\end{equation}
or 
\begin{equation}\label{eqn:duS}
    \frac{\partial \uS}{\partial x^i}\Big|_{\xx(\yy^*)} = \frac{\partial \pi^a}{\partial x^i}\Big|_{\xx(\yy^*)} y^*_a,
\end{equation}
which determines the MaxEnt dependence $\mu:\yy^*\mapsto\xx(\yy^*)$. This also defines the \emph{reduced conjugate entropy}
\begin{equation}
    \dS^*(\yy^*) = -\uS(\xx(\yy^*)) + y^*_a \pi^a(\xx(\yy^*)).
    \label{eqn:dSstar}
\end{equation}
The potential
\begin{equation}\label{eq.dSt*}
    \dSt^*(\xx,\yy^*) \stackrel{def}{=} -\uS(\xx) + y^*_a \pi^a(\xx)
\end{equation}
is called the \emph{reducing potential} \cite{grmela-rate}, and will be used later in the generalization of the KL divergence.

Second, the Legendre transform converts the reduced conjugate variables $\yy^*$ to the reduced state variables $\yy$ by solving
\begin{equation}
    \frac{\partial}{\partial \yy^*}\left( -\dS^*(\yy^*) + y^a y^*_a\right) = 0,
    \label{eqn:dS}    
\end{equation}
which gives the mapping $\yy^*(\yy)$ as well as the reduced entropy 
\begin{equation}
    \dS(\yy) = -\dS^*(\yy^*(\yy)) + y^a y^*_a(\yy).
\end{equation}
Altogether, the MaxEnt procedure gives the functions $\xx(\yy^*)$, $\yy^*(\yy)$, the reduced conjugate entropy $\dS^*(\yy^*)$, and the reduced entropy $\dS(\yy)$. Note that the reduced entropy can also be obtained by inserting the mapping $\xx(\yy^*(\yy))$ into the original entropy $\uS(\xx)$, that is $\dS(\yy) = \uS(\xx(\yy^*(\yy)))$. 

Several other formulas related with MaxEnt and useful later in this manuscript can be found in Appendix \ref{sec.MaxEnt.detail}.

\paragraph{Application to probability distributions.}
MaxEnt can be used to find the least biased estimate of a probability distribution $\pp$, playing the role of the detailed variables, based on the knowledge of some moments of the distribution $y^a = \sum_i \pi^a_i p_i$ and the normalization $\nu=\sum_i p_i=1$ (reduced variables). 
In the case of Shannon entropy \eqref{eq.shannon}, the distribution estimate $\pp$ can be calculated as a function of the reduced conjugate variables $\yy^*$ (or Lagrange multipliers) as
\begin{equation}\label{eq.maxent.pi}
    \frac{\partial}{\partial \pp}\underbrace{\left( -\uS(\pp) + y^*_a \pi^a(\pp) + \nu^*\sum_i p_i \right)}_{=\dSt^*(\pp,\yy^*)} = 0 
    \quad\Rightarrow\quad
    p_i(\yy^*) = \frac{1}{\alpha} e^{-y^*_a \pi^a_i/k_B - \nu^*/k_B - 1}. 
\end{equation}
The dependence $p_i(\yy)$ is then obtained either by solving for the Lagrange multipliers or by the Legendre transformation from $\yy^*$ to $\yy$, see \cite{pkg}.
When we carry out the Legendre transformation, the lower conjugate entropy is obtained from the reduction potential as
\begin{align}
    \dS^*(\yy^*) = \dSt^*(\pp(\yy^*),\yy^*) =& 
    k_B \sum_i\frac{1}{\alpha} e^{-y^*_a \pi^a_i/k_B - \nu^*/k_B -1}\left(-y^*_a \pi^a_i/k_B - \nu^*/k_B -1\right) +\nonumber\\
    &+ \sum_i y^*_a \pi^a_i \frac{1}{\alpha} e^{-y^*_a \pi^a_i/k_B - \nu^*/k_B -1} 
    + \sum_i\nu^* \frac{1}{\alpha} e^{-y^*_a \pi^a_i/k_B - \nu^*/k_B -1}\nonumber\\
    &= -k_B \sum_i \frac{1}{\alpha} e^{-y^*_a \pi^a_i/k_B - \nu^*/k_B -1},
\end{align}
and thus $\dS^*(\yy^*)|_{\nu=1}=-k_B$. The reduced entropy $\dS(\yy)$ can be then obtained by the Legendre transformation, which needs a concrete choice of the moments $\pi^a_i$. 

\paragraph{Concrete examples.}
The simplest case is when only the normalization is known, $\pi^a_i=0$, where the final Legendre transformation gives $\nu = \sum_i \alpha^{-1} e^{-\nu^*/k_B -1}$, or $\nu^* = -k_B \ln(\alpha \nu/W) - k_B$ where $W$ is the number of possible states, $i=1\dots W$. The reduced entropy is then 
\begin{align}
    \dS(\nu) =& -\dS^*(\nu^*(\nu)) + \nu \nu^*(\nu) = k_B\nu + \nu(-k_B \ln(\alpha \nu/W) - k_B)\nonumber\\
    =& k_B \nu \ln(W/(\alpha \nu))
    =|_{\nu=1,\alpha=1} k_B \ln(W),
\end{align}
which is the Boltzmann microcanonical entropy. If we choose the energy as the only moment, $\pi^1_i = E_i$, we obtain the Boltzmann canonical entropy, and if we choose both the energy and the number of particles as moments, we obtain the Boltzmann grand-canonical entropy \cite{landau5,pkg}.

\subsection{Kullback--Leibler discrepancy: A generalization of Kullback–Leibler divergence}
We now introduce a generalization of the KL divergence that measures the information distance between elements $\xx$ and $\yy$ belonging to two different manifolds related by the MaxEnt procedure. The key definition is as follows.

Given a detailed entropy $\uS(\xx)$, a projection $\pi:\xx\mapsto\yy$, and the reduced entropy $\dS(\yy)$ obtained from $\uS$ by MaxEnt, the \emph{MaxEnt Kullback--Leibler discrepancy} is
\begin{equation}
D^M_{KL}(\xx||\yy) \stackrel{def}{=} D^{M*}_{KL}(\xx || \yy^*)|_{\yy^*(\yy)}
\end{equation}
where the \emph{conjugate MaxEnt KL discrepancy} is
\begin{equation}\label{eqn:DKLM}
    D^{M*}_{KL}(\xx || \yy^*) = \frac{1}{k_B} \left( \dSt^*(\xx,\yy^*) - \dS^*(\yy^*) \right),
\end{equation}
with $\dSt^*(\xx,\yy^*)$ the reducing potential \eqref{eq.dSt*} and $\dS^*(\yy^*)$ the reduced conjugate entropy \eqref{eqn:dSstar}.
This discrepancy is automatically non-negative since $\dSt^*(\xx,\yy^*) \geq \dS^*(\yy^*)$, and it reduces to the standard KL divergence when the Shannon entropy is used as $\uS$ (shown below), while also being applicable to other entropies such as Tsallis--Havrda--Charv\'{a}t (see Appendix \ref{sec.tsallis}). Moreover, for affine mappings $\pi$, the MaxEnt KL discrepancy is convex in $\xx$ and it becomes equivalent to Bregman divergence \cite{bregman}, while keeping the possibility of nonlinear mappings $\pi$. We use the term \emph{discrepancy} rather than divergence because $\xx$ and $\yy^*$ need not be elements of the same manifold. 

The defintion of the MaxEnt KL discrepancy is not arbitrary, as entropy is the unique measure of uncertainty that satisfies the Shannon postulates \cite{shannon, jizba-maxent}, and thus it provides the least biased estimate of the information loss when we reduce the description from $\xx$ to $\yy$. Moreover, the MaxEnt KL discrepancy has several desirable properties, such as non-negativity and recovery of the standard KL divergence in the case of Shannon entropy, which we now show.

\paragraph{Non-negativity.}
The non-negativity follows from the Young--Fenchel inequality. For an entropy $\uS(\pp)$ as a function of a probability distribution, the conjugate entropy is $\uS^*(\pp^*) = \inf_{\pp}(-\uS(\pp) + p_i p^*_i)$, and
\begin{equation}
    \uS(\pp) + \uS^*(\pp^*) \leq p_i p^*_i.
\end{equation}
For a second distribution $\qq$ with conjugate $\qq^* = \frac{\partial \uS}{\partial \qq}$, this implies
\begin{equation}
    -\uS(\pp) + p_i q^*_i -\uS^*(\qq^*) \geq  0, 
\end{equation}
which has the form of the MaxEnt KL discrepancy \eqref{eqn:DKLM}.

\paragraph{Recovery of the standard KL divergence.}
In the case of Shannon entropy with $\alpha=1$, the MaxEnt KL discrepancy becomes
\begin{equation}
    D^{M*}_{KL}(\pp || \qq) = \frac{1}{k_B} \left( k_B \sum_i p_i \ln(p_i) + p_i q^*_i +k_B\right),
\end{equation}
and for $\qq^*_i = -k_B \ln(q_i) - k_B$ it simplifies to the usual KL divergence \eqref{eq.KL}.
In Appendix \ref{sec.tsallis}, we show how the MaxEnt KL discrepancy works in the case of Tsallis-Havrda-Charvát entropy \cite{tsallis,havrda-charvat,jizba-maxent}.

\subsection{Fisher information matrix}
The \emph{Fisher information matrix} \cite{fisher,rao} is a measure of the amount of information that a distribution $\pp$ carries about an unknown parameter $\yy$,
\begin{equation}\label{eqn:fisher}
    I_{ab}(\pp, \yy) = \sum_i \frac{\partial \ln p_i}{\partial y^a} \frac{\partial \ln p_i}{\partial y^b} p_i = -\sum_i \frac{\partial^2 \ln p_i}{\partial y^a \partial y^b} p_i,
\end{equation}
and it can be seen as the Hessian of the KL divergence with respect to the parameters $\yy$. 

Similarly, the \emph{MaxEnt Fisher information matrix} can be defined as the pullback of the Hessian with respect to the detailed variables $\xx$,
\begin{equation}\label{eqn:MaxEntFisher}
    I^M_{ab}(\yy) = -\mu^*\left(d^2\uS\right)_{ab} = -\frac{\partial x^i}{\partial y^a}\frac{\partial^2 \uS}{\partial x^i \partial x^j}\Big|_{\xx(\yy)}\frac{\partial x^j}{\partial y^b}
\end{equation}
because
\begin{equation}\label{eq.dSt2}
    \dSt^*(\xx(\yy^*) + d\xx,\yy^*) = \dS^*(\yy^*) + \underbrace{\left(-\frac{\partial \uS}{\partial x^i}\Big|_{\xx(\yy^*)} + y^*_a \frac{\partial \pi^a}{\partial x^i}\Big|_{\xx(\yy^*)}\right)}_{=0} dx^i - \frac{1}{2} \underbrace{\frac{\partial^2 \uS}{\partial x^i \partial x^j}\Big|_{\xx(\yy^*)} d x^i d x^j}_{d^2\uS} + \mathcal{O}(d\xx^3).
\end{equation}
In the case of Shannon entropy $\uS(\pp)$, the MaxEnt Fisher information matrix coincides with the standard Fisher information matrix \eqref{eqn:fisher}. The MaxEnt Fisher information matrix \eqref{eqn:MaxEntFisher} is a generalization of the standard Fisher information matrix \eqref{eqn:fisher} because it does not rely on the Shannon form of the entropy.

The MaxEnt Fisher information matrix can be also expressed in terms of the conjugate variables $\yy^*$ as
\begin{equation}
    I^{M*}_{ab}(\yy^*) = -\frac{\partial x^i}{\partial y^*_a}\frac{\partial^2 \uS}{\partial x^i \partial x^j}\Big|_{\xx(\yy*)}\frac{\partial x^j}{\partial y^*_b},
\end{equation}
which will be useful later in this manuscript.

\section{Lack-of-fit reduction}\label{sec.lof}
In this section, we provide a variant of the lack-of-fit reduction in non-equilibrium thermodynamics that is based on the Onsager-Machlup path-integral formalism \cite{om}, generalizing works of Bruce Turkington \cite{turkington,Turkington-turbulence,Turkington-point-vortex,Turkington-fluctuations}, Richard Kleeman \cite{kleeman}, and ours \cite{JSP2020,lof2024}. 

The lack-of-fit method reduces a detailed description of a physical system to a less detailed description by taking into account only some degrees of freedom. The basic idea is to minimize the \emph{discrepancy} (lack of fit) between the detailed and reduced evolution equations near the \emph{MaxEnt submanifold} (the image of the MaxEnt mapping from the reduced state space to the detailed space). The reduced dynamics then contains dissipative terms that are derived from the detailed evolution. In particular, even purely Hamiltonian systems can be reduced to a dissipative dynamics, and the method does not require any fitting parameters. In the present manuscript, we formulate the lack-of-fit reduction in the context of the KL discrepancy introduced in the preceding chapter.%the path-integral formalism \cite{kleeman}.

\subsection{Information-geometric formulation}\label{sec.lof.ig}
Suppose now that some detailed variables $\xx$ are reduced to a set of reduced variables $\yy$ by a projection $\pi:\xx\mapsto\yy$. If one knows only the reduced variables, the least biased estimate of the detailed variables can be expressed as functions of the reduced variables (or their conjugates) as the MaxEnt mapping $\xx(\yy^*)$. We assume that the detailed system is such that it stays close to the MaxEnt submanifold, and if it deviates from the submanifold, it falls back within a typical time much shorter than $(\Delta t)_y$, as in \cite{gk}. 

\paragraph{Lack-of-fit Lagrangian.}
The lack-of-fit Lagrangian is defined as the MaxEnt KL discrepancy between the detailed evolution of $\xx$ and the MaxEnt estimate of the evolution based on $\yy^*$, divided by the time-scale of the reduced variables $(\Delta t)_y$,
\begin{equation}
\mathcal{L}_{\text{LoF}}(\xx,\yy^*)
=\frac{1}{(\Delta t)_y}\left(\dSt^*(\xx,\yy^*)-\dS^*(\yy^*)\right),
\label{eqn:Lagrangian}
\end{equation}
and it represents the cost function for the lack-of-fit reduction. The total lack-of-fit discrepancy accumulated over a time interval is then obtained by summing this cost over all intermediate points:
\begin{align}
\Delta_{\text{LoF}} =& \sum_\alpha \mathcal{L}_{\mathrm{LoF}}(\xx(t_\alpha),\yy^*(t_\alpha)) (\Delta t)_y = \sum_\alpha D^{M*}_{KL}(\xx(t_\alpha) || \yy^*(t_\alpha)),
\label{eqn:Delta_LOF}
\end{align}
where we used $\mathcal{L}_{\mathrm{LoF}}=\frac{k_B}{(\Delta t)_y}D^{M*}_{KL}(\xx\|\yy^*)$.
The factor $(\Delta t)_y$ is the time-scale of the reduced variables, which is typically much larger than the time-scale of the detailed variables $(\Delta t)_x$. 

\paragraph{Expansion around MaxEnt.}
Let us simplify the lack-of-fit Lagrangian by expanding $\dSt^*(\xx,\yy^*)$ around the MaxEnt point $\xx(\yy^*)$:
\begin{equation}
\dSt^*(\xx(\yy^*)+\delta\xx,\yy^*)
=\dS^*(\yy^*)+\underbrace{\frac{\partial \dSt^*}{\partial x^i}}_{=0}\delta x^i
+\frac{1}{2}\delta x^i\frac{\partial^2\dSt^*}{\partial x^i\partial x^j}\Big|_{\xx(\yy^*)}\delta x^j+o(\|\delta\xx\|^2),
\end{equation}
where the linear term vanishes by the MaxEnt condition. Using \eqref{eq.dSt2}, this becomes
\begin{equation}
D^{M*}_{KL}(\xx+\delta \xx || \yy^*) \approx -\frac{1}{2k_B}  \delta x^i \frac{\partial^2 \uS}{\partial x^i \partial x^j}\Big|_{\xx(\yy^*)} \delta x^j,
\end{equation}
where $\delta \xx$ is the difference between the exact evolution of $\xx$ and the MaxEnt estimate of the evolution.

\begin{figure}[ht!]
\centering
\includegraphics[scale=0.5]{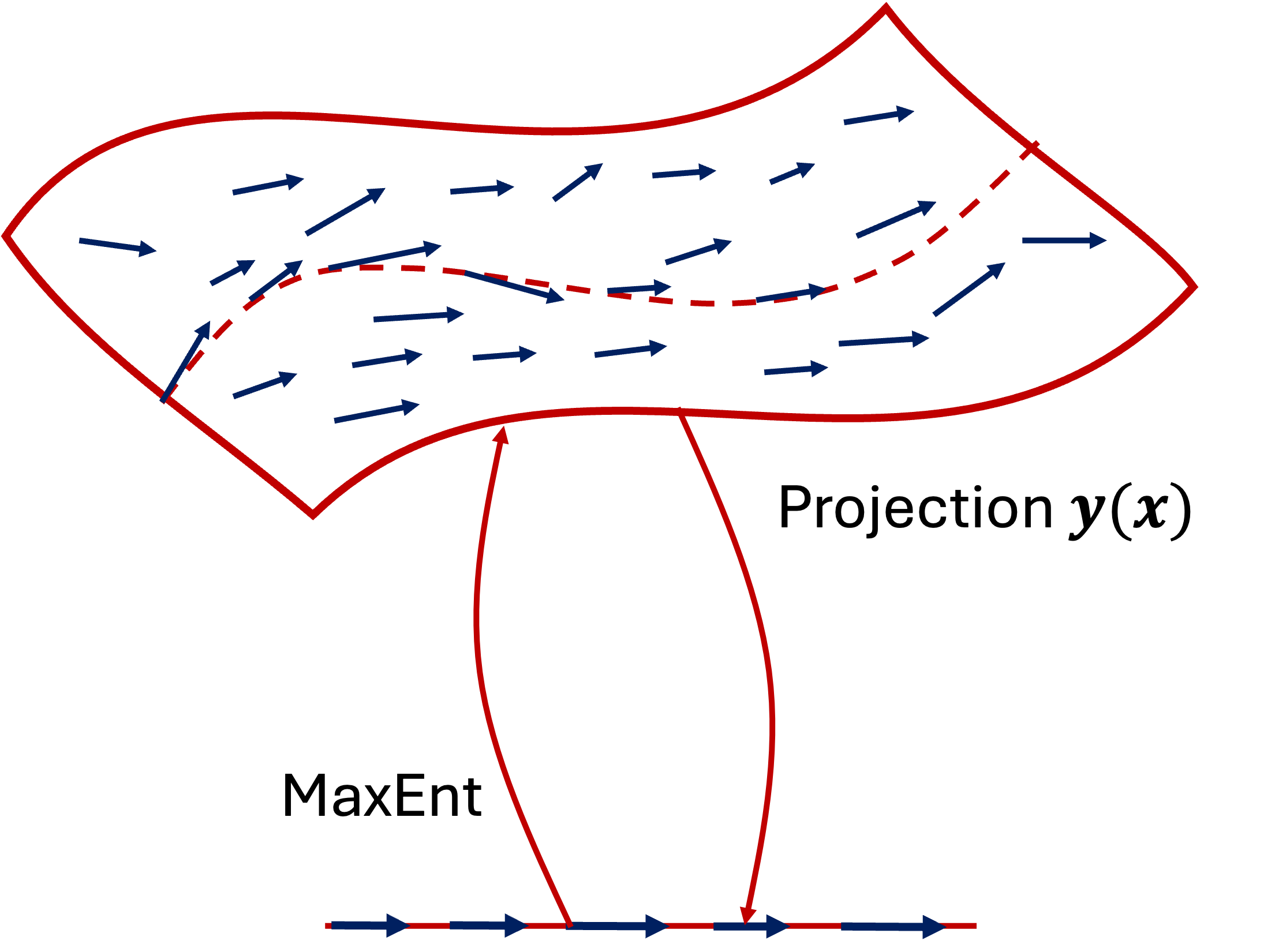}
\caption{The lack-of-fit discrepancy is measured as the difference between the detailed vector field $\dot{\xx}$ in the vicinity of the MaxEnt submanifold and the MaxEnt image of the reduced vector field $\dot{\yy}^*$.}
\label{fig.reduction}
\end{figure}

More precisely, when the system is in state $\xx(\yy^*)$ at time $t$, then at a slightly later time $t+(\Delta t)_x$, the detailed evolution is at $\xx_1 = \xx(\yy^*) + \Delta \xx$. The MaxEnt image $\xx(\yy^*)$, which evolves with the reduced variables, is at $\xx_2 = \xx(\yy^*) + (\Delta \xx)_y$, where $(\Delta \xx)_y \approx \frac{\partial \xx}{\partial \yy^*} \Delta \yy^*$. The expansion of the MaxEnt KL discrepancy with $\delta \xx = \xx_2-\xx_1$ can be then expressed as
\begin{align}
D^{M*}_{KL}(\xx+\delta\xx || \yy^*) \approx & -\frac{1}{2k_B} (x^i_2-x^i_1)\frac{\partial^2 \uS}{\partial x^i \partial x^j}\Big|_{\xx(\yy^*)} (x^j_2-x^j_1) \nonumber\\
=& -\frac{1}{2k_B} (\Delta x^i - (\Delta x^i)_y)\frac{\partial^2 \uS}{\partial x^i \partial x^j}\Big|_{\xx(\yy^*)} (\Delta x^j - (\Delta x^j)_y)\nonumber\\
=& -\frac{(\Delta t)^2_x}{2k_B}
\left(\frac{\Delta x^i}{(\Delta t)_x} - \frac{(\Delta x^i)_y}{(\Delta t)_x}\right)\frac{\partial^2 \uS}{\partial x^i \partial x^j}\Big|_{\xx(\yy^*)} \left(\frac{\Delta x^j}{(\Delta t)_x} - \frac{(\Delta x^j)_y}{(\Delta t)_x}\right).
\end{align}
The information discrepancy \eqref{eqn:Delta_LOF} can then be approximated as
\begin{align}\label{eqn:Delta_LOF_approx}
\Delta_{\text{LoF}} \approx& -\frac{1}{2k_B}\frac{(\Delta t)_x^2}{(\Delta t)_y} \sum_\alpha (\Delta t)_y \left(\frac{\Delta x^i}{(\Delta t)_x} - \frac{(\Delta x^i)_y}{(\Delta t)_x}\right)\frac{\partial^2 \uS}{\partial x^i \partial x^j}\Big|_{\xx(\yy^*)} \left(\frac{\Delta x^j}{(\Delta t)_x} - \frac{(\Delta x^j)_y}{(\Delta t)_x}\right)\nonumber\\
\approx & -\frac{\Delta t}{2k_B}  \int\limits_0^T dt \left(\dot{\xx}^i(\yy^*) - \frac{\partial \xx^i}{\partial \yy^*_a} \dot{\yy}^*_a\right)\frac{\partial^2 \uS}{\partial x^i \partial x^j}\Big|_{\xx(\yy^*)} \left(\dot{\xx}^j(\yy^*) - \frac{\partial \xx^j}{\partial \yy^*_b} \dot{\yy}^*_b\right),
\end{align}
where the time pre-factor is defined as the ratio of the short and slow time-scales multiplied by the short time-scale, that is $\Delta t = \frac{(\Delta t)_x^2}{(\Delta t)_y}$, and the integration is understood as the Riemann sum with uniform time steps $(\Delta t)_y$. Figure \ref{fig.reduction} illustrates the construction of the variational lack-of-fit discrepancy.

The total lack-of-fit discrepancy \eqref{eqn:Delta_LOF_approx} can be seen as an action with the Lagrangian
\begin{equation}
\mathcal{L} (\mathbf{y}^*, \dot{\mathbf{y}}^*)= - \frac{1}{2} \left( R^i \frac{\partial^2 (^{\uparrow}S)}{\partial x^i \partial x^j}\Big|_{\xx(\yy^*)} R^j  \right),
    \label{eqn:Lagrangian_GENERIC}
\end{equation}
where the \emph{residuum} $R^i$ is
\begin{equation}
R^i ( \mathbf{x}(\mathbf{y}^*))=  
^\uparrow L^{ij}(\xx(\yy^*)) \frac{\partial ^\uparrow \overline{E}}{\partial x^j} + \frac{\partial ^{\uparrow}  \Xi}{\partial x^*_i} \Bigg|_{\mathbf{x}^* = \frac{\partial \uS}{\partial {\mathbf{x}}}}
- \frac{\partial x^i}{\partial y^*_a} \dot{y}^*_a,
\label{eqn:Residuum}
\end{equation}
similarly as in \cite{JSP2020,lof2024}. The residuum expresses the difference between the vector field on the MaxEnt submanifold and the image of the reduced vector field obtained by the MaxEnt mapping. Note that when using this quadratic expansion, we assume that the detailed evolution is close to the MaxEnt submanifold.

The \emph{apparent detailed energy} $^\uparrow\overline{E}(\xx) = {^\downarrow}E(\yy(\xx))$, which is present in the residuum, is defined as the projection pullback of the reduced energy
\begin{equation} 
    \dE(\yy) \stackrel{def}{=}\uE(\xx(\yy)).
\end{equation}
The detailed evolution vector field $\XX$ is assumed to be of the GENERIC form \eqref{eq.X}. In particular, it can be purely Hamiltonian without any dissipation potential.

%NOTE: The integration inside the exponential in $W$ in \cite{kleeman} goes to $t$ and it is unclear whether that's a mistake (and the limit should really be T) or if the upper boundary is a variable (?).
\subsection{Lack-of-fit variational principle} 
The reduced path $\yy^*(t)$ is then obtained by minimizing the action
\begin{equation}
\Sigma^*[\mathbf{y}^*(t)] \equiv \int_{\yy^*(t):\,\yy^*(t_0)=\yy^*_0,\,\yy^*(t_1)=\yy^*_1}
 \mathcal{L}_{\text{LoF}} (\mathbf{y}^*(t), \dot{\mathbf{y}}^*(t)) dt.
\label{eqn:DissipationPotential_Turkington}
\end{equation}
As we do not know how long the evolution will take, we have to consider the initial point or the final point and the corresponding times as free. And as we expect the equilibrium to by approached in the long-time limit ($\yy^*(\infty) = \yy^*_{eq}$), we will keep the initial point free while fixing the final point. The extremal action $\Sigma^*_e$ (evaluated along the geodesics) then satisfies the Hamilton--Jacobi equation \cite{GF},
\begin{align}\label{eqn:HJf}
\frac{\partial \Sigma^*_e}{\partial t_0} = \mathcal{H}(\yy^*_0, \pp)|_{\pp=-\frac{\partial \Sigma^*_e}{\partial \yy^*_0}}
\end{align}
where the Hamiltonian is the Legendre transform of the Lagrangian, $\mathcal{H}(\yy,\pp)=\sup_{\dot\yy^*}(-\mathcal{L}+\dot\yy^*\cdot\pp)$. The geodesics are described by the Hamilton equation
\begin{equation}\label{eq.geodesics}
    \dot{\yy}^* = \frac{\partial \mathcal{H}}{\partial \pp}
\end{equation}
with $\pp = -\frac{\partial \Sigma^*_e}{\partial \yy^*}$.

Because the Lagrangian does not explicitly depend on time, the Hamiltonian is a constant of motion. Moreover, since the lack-of-fit discrepancy approaches zero as the system is reaching the equilibrium, the Hamiltonian itself is zero, and thus the extremal action $\Sigma^*_e$ satisfies the stationary Hamilton--Jacobi equation
\begin{equation}\label{eq.HJ.stationary}
\mathcal{H}\left(\yy^*,-\frac{\partial \Sigma^*_e}{\partial \yy^*}\right) = 0,
\end{equation}
which is a PDE for $\Sigma^*_e$ and we solve it later in the concrete examples.

\paragraph{Quadratic form of the Lagrangian and the Hamiltonian.}
Since the Lagrangian \eqref{eqn:Lagrangian_GENERIC} is quadratic in $\dot{\mathbf{y}}^*$, we may write it in the form 
\begin{equation}
\mathcal{L}_{\text{LoF}} (\yy^*, \dot{\mathbf{y}}^*(\dot{\yy})) = \frac{1}{2} \dot{\mathbf{y}}^{*T} g^{-1} \dot{\mathbf{y}}^* + \dot{\mathbf{y}}^* \dXX + \frac{1}{2}||\XX||^2 
\end{equation}
where 
\begin{subequations}
\begin{equation}\label{eq.gab}
g^{-1}: \hskip0.1cm g^{ab} = - \frac{\partial x^i}{\partial y^*_a} \frac{\partial^2 (^{\uparrow}S)}{\partial x^i \partial x^j}\Big|_{\xx(\yy^*)} \frac{\partial x^j}{\partial y^*_b}
,\quad
g_{ab} = -\frac{\partial^2 \dS}{\partial y^a \partial y^b}\Big|_{\yy(\yy^*)}
\end{equation}
is the Fisher information matrix \eqref{eqn:MaxEntFisher} and its inverse (see Appendix \ref{sec.MaxEnt.detail} and Eq. \eqref{eq.d2uS}, assuming that the reduction mapping $\pi$ is affine),
\begin{equation}
\dX^b = \frac{\partial x^j}{\partial y^*_b} \frac{\partial^2 (^{\uparrow}S)}{\partial x^j \partial x^i} \left( ^\uparrow L^{ik}  \frac{\partial ^\uparrow \overline{E}}{\partial x^k} + \frac{\partial ^{\uparrow}  \Xi}{\partial x^*_i} \Bigg|_{\mathbf{x}^* = \frac{\partial \uS}{\partial {\mathbf{x}}}} \right)
=\dL^{bc}(\yy(\yy^*)) \frac{\partial \dE}{\partial y^c}\Bigg|_{\yy(\yy^*)} + \frac{\partial x^j}{\partial y^*_b}\frac{\partial \^uparrow\Xi}{\partial x^*_i}\Bigg|_{\mathbf{x}^* = d\uS(\xx(\yy^*))}
\end{equation}
encodes the detailed evolution projected on the reduced state space, and 
\begin{equation}
||\XX||^2 = - \left( ^\uparrow L^{ik}  \frac{\partial ^\uparrow \overline{E}}{\partial x^k} + \frac{\partial ^{\uparrow}  \Xi}{\partial x^*_i} \Bigg|_{\mathbf{x}^* = \frac{\partial \uS}{\partial {\mathbf{x}}}} \right) \frac{\partial^2 (^{\uparrow}S)}{\partial x^i \partial x^j} \left( ^\uparrow L^{jk}  \frac{\partial ^\uparrow \overline{E}}{\partial x^k} + \frac{\partial ^{\uparrow}  \Xi}{\partial x^*_j} \Bigg|_{\mathbf{x}^* = \frac{\partial \uS}{\partial {\mathbf{x}}}} \right)\Big|_{\xx(\yy^*)}
\end{equation}
measures the magnitude of the detailed evolution vector field. Note that all $\gmetric$, $\dXX$, and $||\XX||^2$ can be functions of $\yy$. 
\end{subequations}

For the quadratic Lagrangian \eqref{eqn:Lagrangian_GENERIC}, the Hamiltonian is
\begin{equation}
\mathcal{H}(\yy^*, \pp) = \frac{1}{2}( \pp - \dXX)^T \gmetric (\pp - \dXX) - \frac{1}{2} ||\XX||^2,
\label{eqn:Hamiltonian}
\end{equation}
and the geodesics are given by Eq. \eqref{eq.geodesics}, which can be also expressed as
\begin{equation}\label{eq.geodesics.explicit}
\dot\yy^* = \gmetric(\pp - \dXX) = -\gmetric\dXX + \gmetric \nabla \Sigma^*_e,
\end{equation}
where $\Sigma^*_e$ is determined by the stationary Hamilton--Jacobi equation \eqref{eq.HJ.stationary}, which for the quadratic Hamiltonian \eqref{eqn:Hamiltonian} becomes
\begin{equation}\label{eq.HJ.explicit}
\frac{1}{2}(\dXX+\nabla\Sigma^*_e)^T \gmetric (\dXX+\nabla\Sigma^*_e) = \frac{1}{2}||\XX||^2.
\end{equation}

\paragraph{Reduced evolution.} This geodesic \eqref{eq.geodesics.explicit} can be also transformed to evolution in the reduced space (see \cite{lof2024} for details), 
\begin{align}\label{eq.reduced}
    \dot{\yy}^a =& \frac{\partial y^a}{\partial y^*_b} \dot{y}^*_b = -g^{ab} \left(-\dX_b -\frac{\partial \Sigma^*_e}{\partial y^*_b}\right)
     = \dL^{ab} \frac{\partial \dE}{\partial y^b} + \frac{\partial}{\partial y^*_a} \left( ^{\uparrow}  \Xi|_{\xx^* = \frac{\partial \uS}{\partial {\mathbf{x}}}(\xx(\yy^*))} + \Sigma^*_e \right),
\end{align}
which means that the reduced evolution has the GENERIC form, provided all the standard properties of the building blocks are satisfied, which we discuss later. The evolution consists of three terms. The first term is the Hamiltonian part, which is the projection of the detailed Hamiltonian evolution on the reduced state space. Note, however, that in general the Jacobi identity for $\dL$ is not guaranteed, unless the expression for $\dL$ does not depend on $\xx$ without having to evaluate $\xx$ on the MaxEnt submanifold. The second term is the gradient flow on the reduced state space driven by the dissipation potential on the detailed space. The third term is the gradient flow driven by the lack of fit between the detailed and reduced vector fields, and this is where irreversibility emerges even out of purely Hamiltonian systems, as illustrated in the examples in Sections \ref{sec.KZ} and \ref{sec.diffusion}.

\paragraph{Tangent formulation.}
The lack-of-fit Lagrangian can be also expressed in terms of the tangent vector $\dot\yy$ instead of the cotangent vector $\dot{\yy}^*$. This is done by using the Legendre transform $\dot{\yy}^* = \frac{\partial \yy}{\partial \yy^*}\dot{\yy}^*$, which gives 
\begin{equation}
\mathcal{L}(\yy,\dot{\yy})= \frac{1}{2} \dot{\yy}^{T} g \dot{\yy} - \dot{\yy} g \dXX + \frac{1}{2}||\XX||^2.
\end{equation}
The corresponding Hamiltonian is then
\begin{equation}
\mathcal{H}(\pp, \yy) = \frac{1}{2}(\gmetric^{-1} \pp + \dXX)^T \gmetric (\gmetric^{-1} \pp + \dXX) - \frac{1}{2} ||\XX||^2,
\label{eqn:Hamiltonian.tangent}
\end{equation}
and the geodesics are given by
\begin{equation}
\dot\yy = \gmetric^{-1} \pp|_{\pp=-\nabla_\yy \Sigma_e} + \dXX = \dXX + \frac{\partial \yy}{\partial \yy^*} \frac{\partial \Sigma_e}{\partial \yy}
=\dXX + \frac{\partial \Sigma_e(\yy(\yy^*))}{\partial \yy^*}, 
\end{equation}
so the tangent action is the pull-back of the cotangent action, $\Sigma_e(\yy) = \Sigma^*_e(\yy^*)$. The Hamilton--Jacobi equation for the tangent action is then
\begin{equation}\frac{1}{2}\left(\dXX + \frac{\partial \Sigma_e}{\partial \yy^*}\right)^T \gmetric \left(\dXX + \frac{\partial \Sigma_e}{\partial \yy^*}\right) = \frac{1}{2}||\XX||^2.
\end{equation}
The geodesics minimizing the tangent action then lead directly to Eq. \eqref{eq.reduced}.

\paragraph{Dissipativeness.} Both the Lagrangian and the Hamiltonian are convex in $\dot{\yy}$ and $\pp$, respectively, which makes the Legendre transform well-defined. If, moreover, the Lagrangian is jointly convex in $\yy$ and $\dot{\yy}$, then lack-of-fit action, which plays the role of a dissipation potential, is also convex (as can be shown by interpolating the action between two geodesics). In the concrete examples in Sections \ref{sec.KZ} and \ref{sec.diffusion}, we show that the lack-of-fit action is indeed convex, which guarantees the dissipativeness of the reduced evolution, without any fitting parameters.

\section{Path-integral formulation of lack-of-fit reduction}\label{sec.path_integral}
In this section, we formulate the lack-of-fit reduction in terms of path integrals, following \cite{kleeman,om,kraaj}. Since the reduced dynamics lacks some information about the detailed dynamics, it can be expected that it will not exactly reproduce the detailed evolution of the reduced variables, $\dot{\yy} \neq \frac{\partial \yy}{\partial \xx} \dot{\xx}$. The detailed (often microscopic) motion that is not seen in the reduced evolution equations can be then modelled as noise on the reduced level of description. The path-integral formalism provides a natural framework for such a stochastic description of the reduced dynamics.

\subsection{Stochasticity}
When a system is described by a Langevin equation, probability of a path can be expressed as the exponential of an action \cite{ventsel}, and the minimum of the action then gives the \emph{most probable path}. 

The path-integral formalism builds upon the \emph{Onsager-Machlup} (OM) \emph{variational principle} \cite{om}. The principle is formulated as a Wiener path integral with the \emph{path amplitudes} $W$ 
\begin{equation}
\mathcal{P}[\yy(t)] = C \exp{\left[- \frac{\Delta t}{k_B} \int\limits_0^T \mathrm{d}t \mathcal{L}\left(\yy, \dot{\yy}\right) \right]},
    \label{eqn:path_amplitudes_W}
\end{equation}
where $\Delta t$ is a characteristic time-scale, $k_B$ is the Boltzmann constant, $\yy(t)$ is a path of some chosen variables, $\mathcal{L}\left(\yy, \dot{\yy}\right)$ is a Lagrangian, and $C$ is a normalization constant. In the OM near-equilibrium case, $\mathcal{L}\left(\yy, \dot{\yy}\right) = (\dot{\yy} - \UU\yy)^T \gmetric (\dot{\yy} - \UU\yy)$, with $\gmetric$ and $\UU$ being constant matrices. The probability $p(\yy,T)$ as a function of the variable $\yy$ at a particular time $T$ is \cite{kleeman}
\begin{equation}
p(\yy_T,T) = C  \int d\yy_0 p(\yy_0) K(\yy_0, \yy_T),
    \label{eqn:path_integral_probab}
\end{equation}
with the kernel
\begin{equation}
K(\yy_0, \yy_T) = C \int_{\yy(0)=\yy_0, \yy(T)=\yy_T} \mathcal{P}[\yy(t)] \mathcal{D}\yy.
    \label{eqn:path_integral}
\end{equation}
The most probable path is then obtained by minimizing the action $\int_0^T \mathcal{L}(\yy,\dot{\yy})\mathrm{d}t$ as $\dot{\yy} = \UU\yy$.

The path integral then becomes dependent on the variables,
\begin{equation}
K(\mathbf{y}_0, \mathbf{y}_T) = \int_{\mathbf{y}(0)=\mathbf{y}_0, \mathbf{y}(T)=\mathbf{y}_T}  C \exp{\left(- \frac{\Delta t}{k_B}    \int\limits_0^T \mathrm{d}t  \mathcal{L} (\mathbf{y}, \dot{\mathbf{y}})\right)}\mathcal{D}\mathbf{y}.
    \label{eqn:path_integral_GENERIC}
\end{equation}

The lack-of-fit reduction provides us with the Lagrangian $\mathcal{L}(\yy, \dot{\yy})$ in \eqref{eqn:Lagrangian_GENERIC}, which can be used in the path-integral formulation \eqref{eqn:path_integral_GENERIC} to obtain probabilities of paths of the reduced variables $\yy$. The path minimizing the action is characterized by the GENERIC evolution equation \eqref{eq.reduced}, which however does not contain any fluctuations. How to extend it to a stochastic evolution that would take into account the unresolved motion of the detailed variables \cite{Ariel2008,FordKacMazur}?

\paragraph{Langevin equation.}
Following \cite{kleeman}, we take the path-integral formulation \eqref{eqn:path_integral_GENERIC}, from which we can reconstruct the corresponding Langevin equation for the reduced state variables $\yy$ 
\begin{equation}\label{eq.Langevin}
    d\mathbf{\yy} = \YY dt + \PPi d\WW 
\end{equation}
where
\begin{equation}
\PPi\PPi^T = \frac{k_B}{\Delta t} \gmetric^{-1}
\end{equation}
and $d\WW$ is a Wiener process. Usually, one starts with the Langevin equation and then derives the path-integral formulation \cite{ventsel}. Here, we proceed in the opposite direction, starting from the path-integral formulation of the lack-of-fit reduction and reconstructing the corresponding Langevin equation. 

Although the reconstruction of the noise is not unique in general due to possible rotations, the matrix $\PPi$ can be determined uniquely if we assume it to be positive semidefinite and symmetric. 

\subsection{Detailed balance}\label{sec.db}
GENERIC can also be seen as the most probable path of stochastic processes obeying a large-deviation principle \cite{kraaj,hco-jnet2020-I,hco-jnet2020-II}. \emph{Microscopic reversibility} means that the probability of a path is equal to the probability of the time-reversed path, and from the physical point of view, it means that the microscopic dynamics is reversible. 

This microscopic reversibility is expressed as the \emph{fluctuation symmetry}, which can be seen as a generalization of the detailed balance condition \cite{boltzmann},
\begin{equation}
    S(\gamma(0)) - \int_0^T  \mathcal{L}(\gamma(t),\dot{\gamma}(t)) dt = S(\gamma(T)) - \int_0^T  \mathcal{L}(\Theta\gamma(t),\Theta\dot{\gamma}(t)) dt,
\end{equation}
where $\gamma(t)$ is a path of the state variables, $S$ is the entropy, and $L$ is the Lagrangian in the large-deviation principle \cite{kraaj}. In the infinitesimal form, the fluctuation symmetry reads
\begin{equation}
    \mathcal{L}(\Theta \yy,\Theta \dot{\yy}) - \mathcal{L}(\yy,\dot{\yy}) = -\frac{\partial S}{\partial y^i} \dot{y}^i,
\end{equation}
which reduces to $\mathcal{L}(\yy,\dot{\yy}) - \mathcal{L}(\yy,-\dot{\yy}) = -\frac{\partial S}{\partial y^i} \dot{y}^i$ when all variables are time-even.

Within the lack-of-fit reduction, using the Onsager--Machlup form $\mathcal{L}=\frac{1}{2}(\dot{\yy}-\YY)^T\gmetric(\dot{\yy}-\YY)$ and the time-reversal decomposition of $\YY$, the antisymmetric part becomes
\begin{equation}\label{eq.lof.db}
    \mathcal{L}(\Theta \yy, \Theta \dot{\yy}) - \mathcal{L}(\yy, \dot{\yy})
    = 2 \left(\dot{y}^a - Y_{\mathrm{rev}}^a\right)\left(g_{ab} Y_{\mathrm{irr}}^b\right),
\end{equation}
with
$Y_{\mathrm{rev}}^a = \dL^{ab} \frac{\partial \dE}{\partial y^b} + g^{ab}\frac{\partial \Sigma^{\textrm{odd}}_e}{\partial y^b}$
and
$Y_{\mathrm{irr}}^a = \Delta_{\textrm{irr}} \pi^a + g^{ab}\frac{\partial \Sigma^{\textrm{even}}_e}{\partial y^b}$,
where $\Sigma_e^{\textrm{even/odd}}$ are the time-even/odd parts under $\Theta$. For details of the calculation see Appendix \ref{appen:detailbalance}.

\section{Kac–Zwanzig example}\label{sec.KZ}
Let us now illustrate the general lack-of-fit reduction on the Kac-Zwanzig model \cite{FordKacMazur,zwanzig,Stuart1999}. The Kac--Zwanzig model is a classical Hamiltonian model of friction that exhibits effectively irreversible evolution of a large particle interacting with many small particles. The reduction of the Kac--Zwanzig model can be calculated analytically for some specific distributions of the frequencies of the small particles (connected via springs to the large particle) \cite{Ariel2008,Kupferman}. However, there is no general analytical theory available for arbitrary distribution of the frequencies, that would not eventually rely on the measurement of the Green-Kubo relations and fluctuations of the detailed dynamics \cite{kubo,zwanzig,grabert,espanol-plateau}. The lack-of-fit reduction provides an alternative way to obtain the reduced evolution for a general distribution of the spring constants.

\subsection{The Kac–Zwanzig model}
The Kac--Zwanzig model is one of the classical Hamiltonian models of friction \cite{FordKacMazur}. It consists of a large particle of mass $M$ moving in a potential $V(\QQ)$ and interacting with $N$ small particles of mass $m_i$ ($i=1, \ldots, N$) via harmonic springs. Under certain conditions, the large particle exhibits effectively irreversible evolution, despite the overall system being reversible. 

It can be shown that the evolution of the large particle is governed by a stochastic differential equation, when the number of small particles goes to infinity \cite{Stuart1999}. Here, we use the Kac–Zwanzig model to illustrate a new general theory of reduction in non-equilibrium thermodynamics.

The Hamiltonian function of the system is
\begin{equation}
H = \frac{\PP^2}{2M}+ V(\QQ) + \sum\limits_{i=1}^N \left[ \frac{\pp_i^2 }{2 m_i} + \frac{\gamma}{2N} \left(\qq_i - \QQ \right)^2 \right],
\label{eqn:HKaz}    
\end{equation}
where $\QQ$ and $\PP$ are the position and momentum of the large particle, $\qq_i$ and $\pp_i$ are the positions and momenta of the small particles, and $\gamma/N$ is the spring constant. The equations of motion implied by this Hamiltonian are 
\begin{subequations}\label{eq.KZ}
\begin{align}
\begin{aligned}
\dot{\QQ} &= \frac{\partial H}{\partial \PP} \\
\dot{\PP} &= -\frac{\partial H}{\partial \QQ}
\end{aligned}
\qquad
\begin{aligned}
\dot{\qq_i} &= \frac{\partial H}{\partial \pp_i} \\
\dot{\pp_i} &= -\frac{\partial H}{\partial \qq_i}.
\end{aligned}
\end{align}
\end{subequations}

Before proceeding to the lack-of-fit reduction itself, let us recall the reduction of the Kac–Zwanzig model based on the projection operator method \cite{zwanzig,hco}.

\subsection{Reduction by the Projection Operator Method}
Using the \emph{projection operators} technique for the Hamiltonian equations, the position and momentum of the large particle can be shown to satisfy the following integro-differential equations \cite{zwanzig}:
\begin{align}
\dot{\QQ} = \frac{\PP}{M} ,& \hskip0.5cm \dot{\PP} = -\frac{\partial V}{\partial \QQ} - \int\limits_0^t \Gamma(s) \frac{\PP (t-s)}{M} \textrm{d}s + \mathbf{F}(t) \\
 \label{eq.Q.PO}& M \ddot{\QQ}  + \frac{\partial V}{\partial \QQ} + \int\limits_0^t \Gamma(s) \dot{\QQ}(t-s) \textrm{d}s = \mathbf{F}(t)
\end{align}
with the memory kernel $\Gamma(s)$ and the force $\mathbf{F}(t)$:
\begin{align*}
\Gamma(t) & = - \Theta(t) \frac{\gamma}{N} \sum\limits_j \cos{(\omega_j t)} = - \Theta(t)  \frac{2}{\pi }\int\limits_0^{\infty} \textrm{d}\omega' \frac{J (\omega')}{\omega'} \cos{( \omega' t)}, \\
\mathbf{F}(t) & =  \frac{\gamma}{N} \sum\limits_j \left[(\qq_j(0)-\QQ(0)) \cos{(\omega_j t)} + \frac{\pp_j(0)}{\omega_j m_j} \sin{( \omega_jt)}\right],
\end{align*}
where the spectral density is defined as
\begin{equation*}
J (\omega) = \frac{\pi }{2} \frac{\gamma}{N}\sum\limits_j \omega_j \delta(\omega - \omega_j).
\end{equation*}

% \noindent The fourier transform of the memory kernel:

% \begin{equation*}
% \tilde{\Gamma}(\omega) = \lim_{\varepsilon \to 0^+} \frac{- i \omega \gamma}{N M} \sum\limits_j \frac{1}{\omega_j^2 - \omega^2 - i \omega \varepsilon} = \lim_{\varepsilon \to 0^+} \frac{- i \omega}{M} \frac{2}{\pi} \int\limits_0^{\infty} \textrm{d}\omega' \frac{J (\omega')}{\omega'} \frac{1}{\omega'^2 - \omega^2 - i \omega \varepsilon},
% \end{equation*}

% \noindent with the spectral density

% \begin{equation*}
% J (\omega) = \frac{\pi }{2} \frac{\gamma}{N}\sum\limits_j \omega_j \delta(\omega - \omega_j) 
% \end{equation*}

% \noindent and the Laplace transform:

% \begin{equation*}
% \hat{\Gamma}(z) =  \frac{z \gamma}{N M} \sum\limits_j \frac{1}{\omega_j^2 - z^2} =  \frac{z}{M} \frac{2}{\pi} \int\limits_0^{\infty} \textrm{d}\omega' \frac{J (\omega')}{\omega'} \frac{1}{\omega'^2 - z^2}.
% \end{equation*}

\paragraph{Markovian approximation.}
In the Markovian approximation, where the large particle is governed by a pair of ordinary differential equations, the \emph{memory kernel} is assumed to be approximated by a delta function, 
\begin{equation}
    \Gamma(t) \approx \delta(t) \underbrace{\int_0^\infty \Gamma(s)ds}_{\stackrel{def}{=} \gamma_{theor}}.
\end{equation} 
Assuming, moreover, that the force in Equation \eqref{eq.Q.PO} is negligible (taking the average over the ensemble of small particles), the effective equation of motion of the large particle simplifies to 
\begin{equation}\label{eq.Q.Markov}
    M \ddot{\QQ} + \frac{\partial V}{\partial \QQ} + \gamma_{theor} \dot{\QQ} = 0,
\end{equation}
where $\gamma_{theor}$ is the theoretical friction coefficient. This equation describes the motion of the large particle within potential field $V(\QQ)$ and with the damping coefficient $\gamma_{theor}$.

\paragraph{Spectral densities.}
In order to calculate the friction coefficient, we assume that $N$ is large enough so that the spectral density $J(\omega)$ can be approximated by a continuous function. 
We shall study two particular cases:
\begin{enumerate}
    \item The distribution of $\omega^2_j$ is uniform between $(0, \omega^2_{max})$. The spectral density and the friction coefficient are given by
    \begin{subequations}\label{eq.gamma1}
\begin{align}
 J (\omega) =& \frac{\pi }{2} \frac{2 \gamma }{\omega^2_{max}} \omega^2   \hskip0.3cm 0 \leq \omega \leq \omega_{max}, \\
\Gamma(t) =&  - \Theta(t) \frac{2 \gamma }{\omega^2_{max}}  \left[ \frac{\omega_{max} \sin(\omega_{max}t)}{t} + \frac{\cos(\omega_{max}t) - 1}{t^2} \right] ,\\
  \gamma_{theor} = &  \hskip0.3cm  0.
\end{align}
\end{subequations}

    \item The distribution of $\omega_j$ is uniform between $(0, \omega_{max})$:
\begin{subequations}\label{eq.gamma2}
\begin{align}
 J (\omega) =& \frac{\pi }{2} \frac{\gamma}{\omega_{max}} \omega  \hskip0.3cm 0 \leq \omega \leq \omega_{max} ,\\
 \Gamma(t) =&  - \Theta(t) \frac{\gamma }{\omega_{max}}  \frac{\sin(\omega_{max}t)}{t} ,\\
  \gamma_{theor} = &\gamma \int\limits_0^{\infty} \frac{\sin(\omega_{max}t)}{\omega_{max} t} \textrm{d}t =  \frac{\gamma \pi}{2 \omega_{max}}  .
\end{align}
\end{subequations}
\end{enumerate}
By comparing Equations \eqref{eq.gamma1} and \eqref{eq.gamma2}, it can be seen that even the sole presence of the friction term in the Markovian limit depends on the \emph{spectral density} of the small particles. While Equation \eqref{eq.Q.Markov} will serve as a benchmark for the numerical simulations for the second distribution ($\omega_j$ uniform), we will show that in the case of the first distribution ($\omega^2_j$ uniform) the effective evolution is damped despite the absence of the friction term in the Markovian limit ($\gamma_{theor} = 0$). The lack-of-fit reduction, on the other hand, will be able to reproduce friction even in the absence of the friction term in the Markovian limit, and stay compatible with the direct numerical simulation.

\subsection{Lack-of-fit reduction}
Let us now apply the lack-of-fit reduction to the Kac–Zwanzig model with the quadratic potential $V(\QQ) = \frac{1}{2}\alpha \QQ^2$. Although the results will be similar to those in \cite{lof2024}, here we use a different formulation of the reduction method, presented in Section \ref{sec.lof}, and we add a stochastic generalization as presented in Section \ref{sec.path_integral}. 

\subsubsection{Formulation of the Lack-of-fit Lagrangian}
The first step is to choose the reduced state variables. While the detailed variables are given by the distribution function of all particles $f(\QQ, \PP, \qq_1, \pp_1, \ldots, \qq_N, \pp_N)$, which follows the Liouville equation, the reduced state variables are then chosen as the following averages,
\begin{subequations}
    \begin{align}
    \bar{\QQ} =& \frac{1}{N!}\int\int\dots\int f(\QQ, \PP, \qq_1, \pp_1, \ldots, \qq_N, \pp_N) \hskip0.1cm \QQ \hskip0.1cm  \textrm{d}\QQ \textrm{d}\PP \textrm{d}\qq_1 \textrm{d}\pp_1 \ldots \textrm{d}\qq_N \textrm{d}\pp_N ,\\
    \bar{\PP} =& \frac{1}{N!}\int\int\dots\int f(\QQ, \PP, \qq_1, \pp_1, \ldots, \qq_N, \pp_N) \hskip0.1cm  \PP \hskip0.1cm  \textrm{d}\QQ \textrm{d}\PP \textrm{d}\qq_1 \textrm{d}\pp_1 \ldots \textrm{d}\qq_N \textrm{d}\pp_N ,\\
    \bar{s} =& \frac{1}{N!}\int\int\dots\int f(\QQ, \PP, \qq_1, \pp_1, \ldots, \qq_N, \pp_N) \left(\frac{1}{N}\sum_{i=1}^N \qq_i -\QQ\right) \textrm{d}\QQ \textrm{d}\PP \textrm{d}\qq_1 \textrm{d}\pp_1 \ldots \textrm{d}\qq_N \textrm{d}\pp_N ,\\
    \bar{e} =& \frac{1}{N!}\int\int\dots\int f(\QQ, \PP, \qq_1, \pp_1, \ldots, \qq_N, \pp_N) H(\QQ,\PP,\qq_1,\pp_1,\dots,\qq_N,\pp_N) \textrm{d}\QQ \textrm{d}\PP \textrm{d}\qq_1 \textrm{d}\pp_1 \ldots \textrm{d}\qq_N \textrm{d}\pp_N,
    \end{align}
which represent the average position and momentum of the distinguished particle, the average distance of the small particles from the distinguished particle, and the average energy of the system, respectively. The Lagrangian is then, see \cite{lof2024} for details,
\end{subequations} 
\begin{equation}\label{eqn:Lagrangian_KZ}
\mathcal{L} = \frac{1}{2} (\dot{\mathbf{y}}^*)^T \mathbb{C} \dot{\mathbf{y}}^* + (\dot{\mathbf{y}}^*)^T \mathbb{B}^T  \mathbf{y^*} + \frac{1}{2} \mathbf{y^*}^T\mathbb{A}\mathbf{y^*},
\end{equation}
where $\mathbf{y^*}^T \equiv (Q^*, P^*, s^*, E^*)$ are the conjugate reduced variables, with matrices
\begin{equation}
\mathbb{A}
=
\frac{1}{E^*}
\begin{pmatrix}
\frac{1}{M} & 0 & -\frac{1}{M} & 0\\
0 & \alpha + \gamma &0 & 0 \\
-\frac{1}{M}  & 0  & \frac{1}{M} + \frac{\overline{\omega^2}}{\gamma}   & 0 \\
0 & 0 & 0 & 0
 \end{pmatrix}, \hskip0.3cm
 \mathbb{B}
=
\frac{1}{E^*}
\begin{pmatrix}
0 & 1 & 0 & - \frac{P^*}{E^*}  \\\
-1 & 0 & 1 & \frac{Q^* - s^*}{E^*} \\
0 & -1 & 0 &  \frac{P^*}{E^*} \\
0 & 0 & 0 & 0
 \end{pmatrix},
 \label{eqn:MatrixAB}
\end{equation}
\noindent and
\begin{equation}
\mathbb{C}
=
\frac{1}{E^*}
\begin{pmatrix}
\frac{1}{\alpha} & 0 & 0 &-\frac{Q^*}{E^*\alpha}\\
0 & M & 0 &- \frac{M P^*}{E^*} \\
0 & 0 & \frac{1}{\gamma}  &- \frac{s^*}{E^*\gamma}\\
-\frac{Q^*}{E^*\alpha}& - \frac{M P^*}{E^*} & - \frac{s^*}{E^*\gamma}& \frac{ k_B E^*( N+1 ) + 2\Sigma}{(E^*)^2} 
 \end{pmatrix},
 \label{eqn:MatrixC}
\end{equation}
where $\overline{\omega^2} = \frac{1}{N}\sum_{i=1}^N \omega_i^2$ is the average of the squared frequencies of the small particles and $E^*$ is the inverse temperature (derivative of entropy with respect to energy). To write the Lagrangian as a function of resolved variables, we write $\yy^* = - \mathbb{C}^{-1}\mathbf{y}$ and the Lagrangian then is 

\begin{equation}\label{eqn:Lagrangian_KZ_TN}
\mathcal{L} = \frac{1}{2} (\dot{\mathbf{y}})^T \mathbb{C}^{-1} \dot{\mathbf{y}} + (\dot{\mathbf{y}})^T \mathbb{C}^{-1}\mathbb{B}^T \mathbb{C}^{-1} \mathbf{y} + \frac{1}{2} \mathbf{y}^T\mathbb{C}^{-1}\mathbb{A}\mathbb{C}^{-1}\mathbf{y}.
\end{equation}

In the notation of Section \ref{sec.lof}, Lagrangian \eqref{eqn:Lagrangian_KZ} gives
\begin{equation}
||\XX||^2 = \yy^T \mathbb{C}^{-1} \mathbb{A} \mathbb{C}^{-1} \yy ; \hskip0.2cm \gmetric^{-1} =  \mathbb{C}; \hskip0.2cm \dXX = -\mathbb{B}^T \mathbb{C}^{-1}  \yy .
\end{equation}

In order to close the reduced evolution equations, we need to determine the extremal action $\Sigma_e$ by solving the Hamilton–Jacobi equation \eqref{eqn:HJf}.

\subsubsection{The Hamilton--Jacobi equation}
Stationary Hamilton–Jacobi equation \eqref{eq.HJ.stationary} becomes
\begin{equation}
    \frac{1}{2} \left[\left(\mathbb{C}\frac{\partial \Sigma_e}{ \partial \yy^T} - \mathbb{B}^T\mathbb{C}^{-1} \mathbf{y}  \right)^T\mathbb{C}^{-1} \left(\mathbb{C} \frac{\partial \Sigma_e}{\partial \yy^T} - \mathbb{B}^T\mathbb{C}^{-1} \yy \right) - \mathbf{y}^T  \mathbb{C}^{-1}\mathbb{A}\mathbb{C}^{-1} \yy \right] = 0
\end{equation}
with the condition at final time $\Sigma_e(\mathbf{y}(T))=0$. 
This terminal normalization is the practical way in which, for the present model, we select the dissipative Hamilton--Jacobi branch; in the free-endpoint language it replaces the transversality condition by anchoring the endpoint at equilibrium. Using an Ansatz $\Sigma_e(\mathbf{y},t) = -\frac{1}{2 E^*}\mathbf{y}^{T} \mathbb{C}^{-1} \mathbb{M}(t) \mathbb{C}^{-1} \mathbf{y}$, the solution is found using the Riccati matrix equation
\begin{equation}
 -\frac{1}{(E^*)^2}\mathbb{M}\mathbb{C}^{-1}\mathbb{M} - \frac{1}{E^*}\mathbb{M} \mathbb{C}^{-1} \mathbb{B}^T - \frac{1}{E^*}\mathbb{B} \mathbb{C}^{-1} \mathbb{M}  + \mathbb{A} -  \mathbb{B} \mathbb{C}^{-1}  \mathbb{B}^T = - \frac{1}{E^*}\dot{\mathbb{M}}
\label{eqn:Riccati}
\end{equation}

with $\mathbb{M}(T)=0$. The Riccati equation has both stable and unstable solutions, but only the stable solution is physically relevant as it corresponds to a positive definite dissipation. The stable solution can be written as
\begin{equation}
\mathbb{M}(t) = \left(\mathbb{W}_{21} - \mathbb{W}_{22} e^{-\Lambda (T-t)}\mathbb{W}_{22}^{-1}\mathbb{W}_{21}e^{-\Lambda (T-t)}\right)\left(\mathbb{W}_{11} - \mathbb{W}_{12} e^{-\Lambda (T-t)}\mathbb{W}_{22}^{-1}\mathbb{W}_{21}e^{-\Lambda (T-t)} \right)^{-1},
\label{eqn:mSolNonStat}
\end{equation}
with $\mathbb{W}_{ij}$ determined from 
\begin{equation*}
\mathbb{H}
=
\begin{pmatrix}
\mathbb{C}^{-1} \mathbb{B}^T & -\mathbb{C}^{-1} \\
   \mathbb{B} \mathbb{C}^{-1}  \mathbb{B}^T - \mathbb{A} & -\mathbb{B} \mathbb{C}^{-1} 
 \end{pmatrix}
 \equiv
\begin{pmatrix}
\mathbb{W}_{11} & \mathbb{W}_{12} \\
\mathbb{W}_{21} & \mathbb{W}_{22}
 \end{pmatrix}
 \begin{pmatrix}
\Lambda & 0 \\
0 & -\Lambda
 \end{pmatrix}
 \begin{pmatrix}
\mathbb{W}_{11} & \mathbb{W}_{12} \\
\mathbb{W}_{21} & \mathbb{W}_{22}
 \end{pmatrix}^{-1},
\end{equation*}

see \cite{Kucera}. The existence of a stable solution is conditioned by the existence of a positive diagonal matrix $\textrm{Re}(\Lambda)$. The stationary solution is found as a limit for $t \to \infty$.

\subsubsection{The reduced deterministic dynamics and Langevin equation}
Within the interval $(0,T)$, the zero-cost trajectory of the Onsager-Machlup like Lagrangian is given by the equation \eqref{eq.reduced}: 

\begin{equation}
\dot{\mathbf{y}} = - \mathbb{B}^T \mathbb{C}^{-1} \yy + \frac{1}{E^*}\mathbb{M} \mathbb{C}^{-1} \yy.
\label{eqn:y_evolutionKZ}
\end{equation}
The first term is the reduced Poisson bracket and therefore is fully reversible, whereas the second term has both dissipative and reversible components. The concrete form of the matrix $\mathbb{M}$ will be determined later by asymptotic analysis and numerically.

The Langevin equation for the reduced variables \eqref{eq.Langevin} is given by
\begin{equation}
\Pi
=
\sqrt{\frac{k_B T}{\Delta t}}
\begin{pmatrix}
\frac{1}{\sqrt{\alpha}} & 0 & 0 \\
0 & \sqrt{M} & 0 \\
0 & 0 & \frac{1}{\sqrt{\gamma}} 
 \end{pmatrix}.
\end{equation}
Parameter $\Delta t = \frac{(\Delta t)_x^2}{(\Delta t)_y}$ is calculated as $\Delta t = \frac{dt_{min}^2}{(\Delta t)_y}$ \eqref{eqn:Delta_LOF_approx}, where $dt_{min} = \frac{2\pi}{\omega_{max}}$ is the smallest typical time of the small particles, $\omega_{max}$ is the highest frequency of the small particles, and $(\Delta t)_y = \frac{2\pi}{\Omega_{y, max}}$ is the smallest typical time of the reduced Poisson bracket and as such depends on the choice of variables $\yy$. In other words, from \eqref{eqn:y_evolutionKZ} we can find $\Omega_{y, max}$ as the highest eigenvalue of $ - \mathbb{B}^T \mathbb{C}^{-1}$. Figure \ref{fig:StochasticKleeman} compares the deterministic and stochastic results.

\subsubsection{Asymptotic solution of the Riccati equation}
In the large-time limit, when $\mathbb{M}$ is approximately constant, the Riccati equation \eqref{eqn:Riccati} reduces to an algebraic equation. By introducing non-dimensional parameters $b=\alpha/\gamma$ and $\rho=b\,\overline{\omega^2}/(\alpha/M)$, and performing a dominant-balance analysis for $\rho\gg 1$ (details in Appendix~\ref{appen:riccati}), the unique positive-definite solution is
\begin{subequations}\label{eq.Masymptotic}
    \begin{align}
    m_{11} &= (M\alpha)^{-1/2} \left(\left(\frac{1}{b}-4-8b\right)+8\sqrt{b(1+b)}\right)^{1/2},\\
    m_{12} &= -1-2b+2\sqrt{b(1+b)},\\
    m_{13} &= -(M\alpha)^{-1/2} \left((1+4b)-b\sqrt{b(1+b)}\right)^{1/2}\rho^{-1/2},\\
    m_{22} &= (M\alpha)^{1/2} \sqrt{b(1+b)} \left(\left(\frac{1}{b}-4-8b\right)+8\sqrt{b(1+b)}\right)^{1/2},\\
    m_{23} &= -1,\\
    m_{33} &= (M\alpha)^{-1/2} \rho^{1/2}.
\end{align}
\end{subequations}
For example, with $M=1099$, $\alpha=1$, $b=1$, $\rho=37$, this asymptotic estimate agrees well with the full numerical solution of the Riccati equation (see Appendix~\ref{appen:riccati} for a detailed comparison).

\subsubsection{Numerical results}
In this section, we present the results of the numerical simulations of the Kac–Zwanzig model, similar to those in \cite{stuart-analysis,hald-kupferman,lof2024}. The detailed results are obtained by directly integrating the equations of motion \eqref{eq.KZ} for the detailed variables. The reduced equations of motion are obtained by integrating equations \eqref{eq.reduced} together with the Riccati equation \eqref{eqn:Riccati} in the deterministic case, and the Langevin equation \eqref{eq.Langevin} in the stochastic case.

\begin{figure}[ht] 
\centering
\includegraphics[scale=0.15]{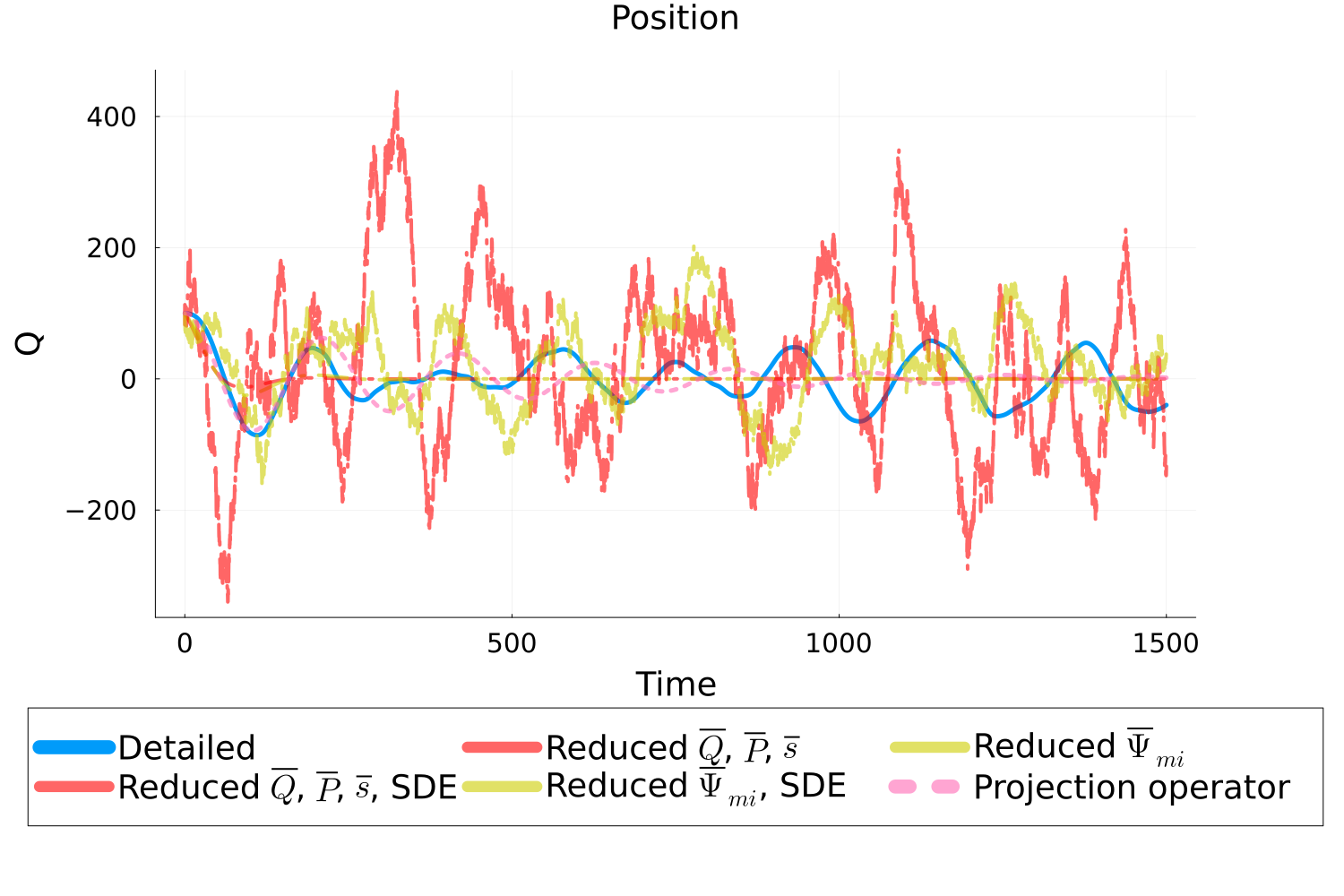}
\includegraphics[scale=0.15]{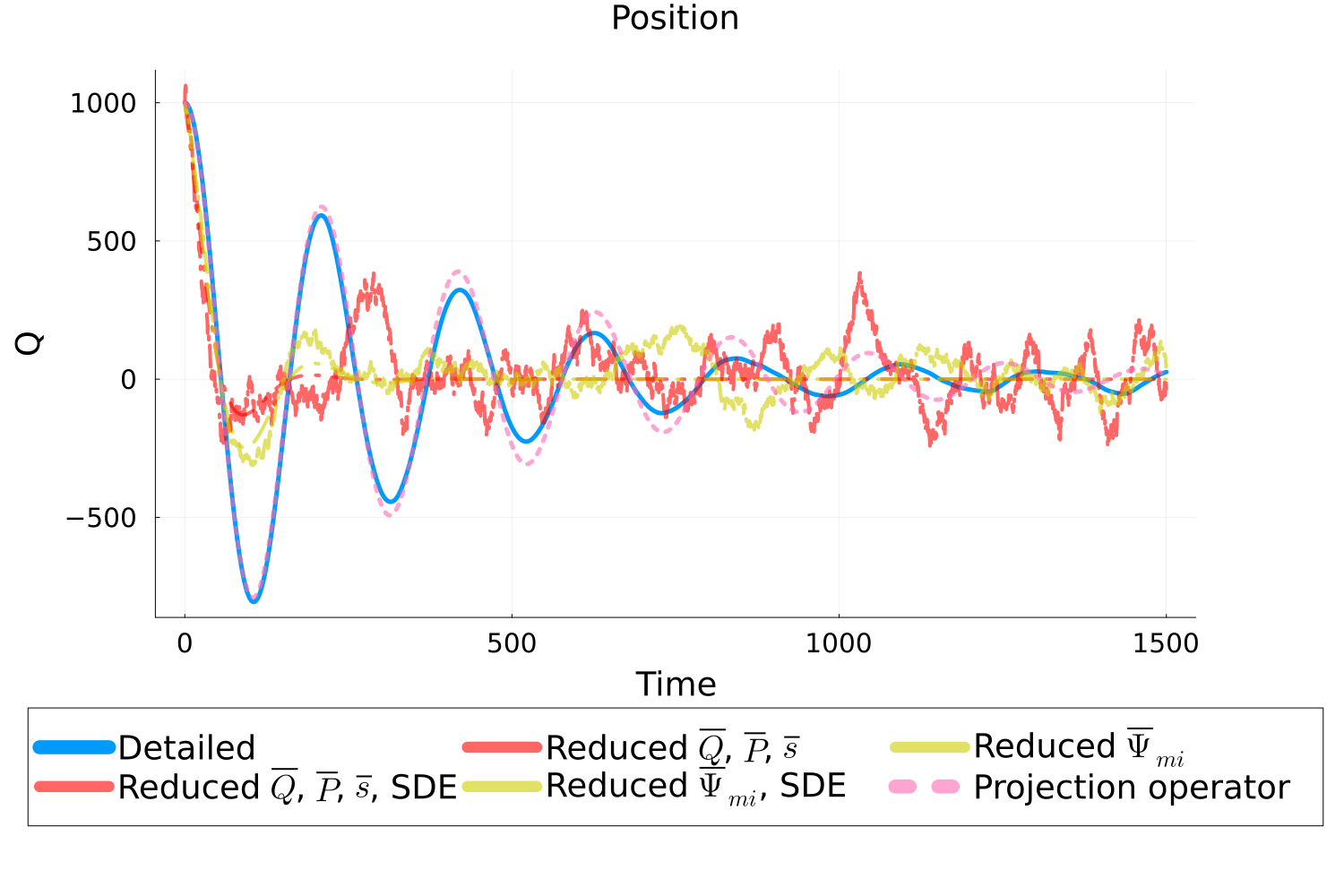}
\caption{Comparison of the deterministic (dotted) and stochastic (full) results for two choices of the reduced state variables ( $\QQ,\PP, s$ (red) \eqref{eqn:y_evolutionKZ} and $\QQ,\PP, s,\Psi_{mi}$ (yellow) \eqref{eq.extra}) and two different initial positions of the big particle (left: $Q_0=100$, right: $Q_0 = 1000$). The reduced dynamics corresponds to the stationary solution of the Hamilton-Jacobi (or Riccati) equation \eqref{eqn:Riccati}, where the effective dissipation potential $\Psi_e(\yy^*)=-\Sigma_e^*(\yy^*)$ is time independent. Parameters of the detailed simulation (blue) were $N=10000$, $\gamma=1.0$, $\alpha = 1.0$, $\omega_i$ sampled from uniformly from the interval $(\sqrt{1.0E-05}, \sqrt{1.0E-01})$, and $M$ was set equal to the total mass of the small particles. In pink is the solution given by the projection operator method \eqref{eq.Q.Markov} that serves as a benchmark for the detailed simulation.}
\label{fig:StochasticKleeman}
\end{figure}

\begin{figure}[ht!]
\centering
\includegraphics[scale=0.25]{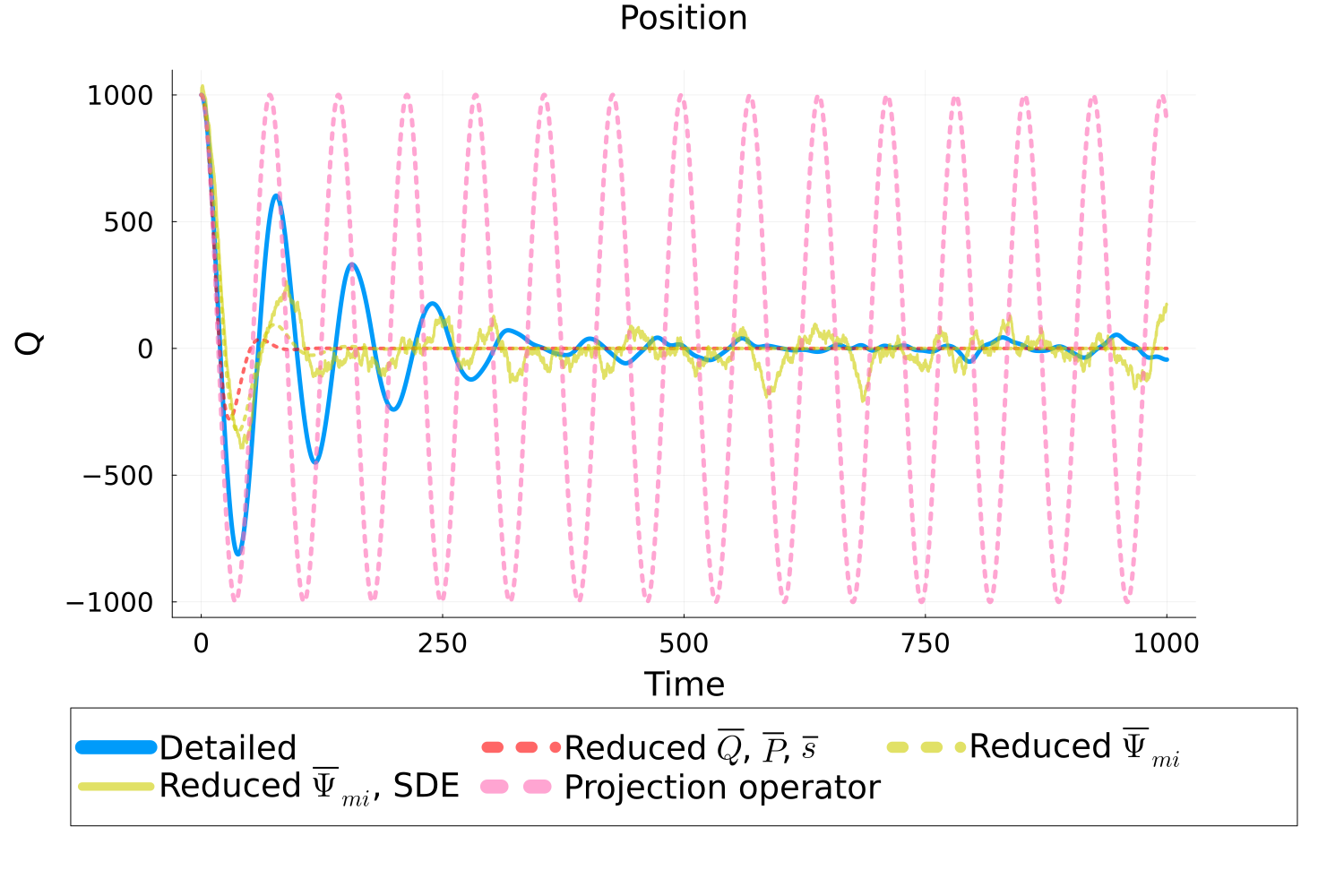}
\caption{Comparison of the deterministic and stochastic results for the initial condition $Q_0 = 1000$ and $\omega_i$ sampled from the uniform distribution $\omega_i^2$ between $10^{-5}$ and $10^{-1}$. The projection operator method gives zero dissipation \eqref{eq.gamma1}. The only difference between this simulation and that in Figure \ref{fig:StochasticKleeman} is the distribution of $\omega_i$.}
\label{fig.omegsq}
\end{figure}

To increase the precision, the simulations are performed for the reduced state variables $\QQ$, $\PP$, and $s$ with extra state variables $\Psi_{mi}$ or $\Psi_\Sigma$, 
\begin{equation}\label{eq.extra}
    \Psi_{mi} = \sum_i \frac{p_i}{N m_i} 
    \quad\text{or}\quad
    \Psi_{\Sigma} = \frac{\sum_i p_i}{N \sum_j m_j},
\end{equation}
see \cite{lof2024} for more details on the calculations.
Figure \ref{fig:StochasticKleeman} shows the trajectories for various initial conditions. It compares two variants of the lack-of-fit reduction with the projection operator method and the direct numerical simulation of the Kac--Zwanzig model. First, the stationary reduction with reduced variables $\QQ$, $\PP$, and $s$, which shows too much damping, is present. The $\Psi_{mi}$ method (with state variables $\QQ,\PP,s,\Psi_{mi}$) is a bit less overdamped. For both choices of variables we also show the stochastic extension.  
The closest agreement between the detailed and reduced dynamics is obtained by the projection operator method. However, for other choices of the distribution $\omega_i$, the projection operator method could even lead to zero dissipation \eqref{eq.gamma1}, as shown in Figure \ref{fig.omegsq}, so we focus on the lack-of-fit methods. 

From the two lack-of-fit methods, the one with $\Psi_{mi}$ gives the better results, whereas the simpler set of state variables, $\QQ$, $\PP$, and $s$ gives the worse result, as expected. The stochastic version is able to reconstruct to some extent the equilibrium fluctuations of the big particle, caused by the motion of the cloud of small particles. This is demonstrated in the simulation with initial condition $Q_0 = 100$ (Figure \ref{fig:StochasticKleeman}).

Section \ref{sec:KLdiv} shows that the KL divergence between the distributions of the detailed and reduced dynamics is the lowest for $\Psi_{mi}$, followed by $\Psi_\Sigma$, and the highest for the simplest set of reduced state variables, which explains the order of precision of those choices \cite{lof2024}.

\subsubsection{KL Divergence}\label{sec:KLdiv}
% Chceme to tu opravdu mít?

Let us suppose that we have two choices of reduced variables $\mathbf{y}$, $\mathbf{z}$ such that their functional derivatives $\frac{\delta \mathbf{y}}{\delta f}$, $\frac{\delta \mathbf{z}}{\delta f}$ are not functions of reduced variables. Their maximum entropy distributions $\tilde{f}(x;\mathbf{y}^*)$, $\tilde{f}(x;\mathbf{z}^*)$ then are of the form:

$$\tilde{f}(x;\mathbf{y}^*) = \frac{1}{\mathcal{Z}(\mathbf{y}^*)} \exp{\left(- \frac{1}{k_B} \frac{\delta \mathbf{y}}{\delta f} \cdot \mathbf{y}^*\right)},$$ with $\mathcal{Z}(\mathbf{y}^*)$ the normalization constant. Their Kullback–Leibler divergence is given as:

$$D_\text{KL}(\tilde{f}(x;\mathbf{y}^*) \parallel \tilde{f}(x;\mathbf{z}^*)) = \int \text{d}\Omega(x) \tilde{f}(x;\mathbf{y}^*) \log\left(\frac{\mathcal{Z}(\mathbf{z}^*)}{\mathcal{Z}(\mathbf{y}^*)}\right)\frac{1}{k_B} \left(\frac{\delta \mathbf{z}}{\delta f} \cdot \mathbf{z}^* - \frac{\delta \mathbf{y}}{\delta f} \cdot \mathbf{y}^*\right).$$

\noindent Final formulas are given in Table \ref{table:KLdiv}, detailed calculations in Appendix \ref{append:KLDivCalc}.
\begin{table}[ht]
\centering
\begin{tabular}{ | c || c | c | c | c | } 
  \hline
   $D_\text{KL}(\downarrow \parallel \rightarrow)$ & $f_{QP}$ & $f_{s}$  & $f_{m_i}$  & $f_{\Sigma}$  \\ 
  \hline \hline
  $f_{QP}$ & 0 & 0 & 0 & 0 \\ 
  \hline
  $f_{s}$ & $\frac{1}{2}\left(\frac{\gamma \bar{s}^2}{k_B T} \right)^2$ & 0 & 0 & 0 \\ 
  \hline
  $f_{m_i}$ & $\frac{1}{2}\left(\frac{\gamma \bar{s}^2 + N\overline{\psi_{m_i}}^2/\overline{\frac{1}{m}} }{k_B T} \right)^2 $ & $\frac{1}{2}\left(\frac{N\overline{\psi_{m_i}}^2}{k_B T \overline{\frac{1}{m}}} \right)^2$ & 0 & $\frac{\Delta_a \Delta_c  \overline{\psi_{m_i}}}{2\overline{\frac{1}{m}}}\left(\frac{N}{k_B T}\right)^2 $ \\ 
  \hline
  $f_{\Sigma}$ & $\frac{1}{2}\left(\frac{\gamma \bar{s}^2 + N\overline{m} N^2 \overline{\psi_{\Sigma}}^2}{k_B T} \right)^2$ & $\frac{1}{2}\left(\frac{N\overline{m} N^2 \overline{\psi_{\Sigma}}^2}{k_B T} \right)^2$ & $ \frac{\Delta_a \Delta_b  N \overline{\psi_{\Sigma}}}{2}\left(\frac{N}{k_B T}\right)^2 $ & 0 \\ 
  \hline
\end{tabular}
\caption{Kullback–Leibler divergence of several distributions. $\Delta_a = \frac{\overline{\psi_{m_i}}^2}{\overline{\frac{1}{m}}} - N^2 \overline{m} \overline{\psi_{\Sigma}}^2 ; \hskip0.3cm \Delta_b = \frac{\overline{\psi_{m_i}}}{\overline{\frac{1}{m}}} - N \overline{m} \overline{\psi_{\Sigma}} ; \hskip0.3cm \Delta_c = \overline{\psi_{m_i}} - N \overline{\psi_{\Sigma}} $}
\label{table:KLdiv}
\end{table}
For our parameters of the Kac--Zwanzig model, the lowest KL divergence is obtained for the $\Psi_{mi}$ distribution, followed by $\Psi_\Sigma$, and the highest for the simplest set of reduced variables $\QQ$, $\PP$, and $s$. This is in agreement with the numerical results \cite{lof2024}.

\section{Diffusion}\label{sec.diffusion}
In this section, we apply the lack-of-fit reduction to a second, physically distinct setting: the emergence of diffusion from the Vlasov kinetic equation. This example serves two purposes. First, it demonstrates that the Hamilton--Jacobi equation for the dissipation potential can be solved in a field-theoretic (infinite-dimensional) context, not only for finite-dimensional ODE systems such as the Kac--Zwanzig model. Second, it reveals a physically important subtlety: for an ideal gas, the lack-of-fit reduction produces a dissipation potential involving a \emph{nonlocal} operator, so that the reduced evolution cannot be written as a diffusion equation. When short-range interparticle interactions are introduced, however, the operator becomes local at short wavelengths and the classical diffusion equation is recovered, with a diffusion coefficient that has the correct temperature scaling.

\subsection{Detailed level: Vlasov equation}
Consider a system of $N$ classical particles with positions $\mathbf{r}^i$ and momenta $\mathbf{p}_i$ in a volume $V$, described by the one-particle distribution function $f(t,\rr,\pp)$. For non-interacting particles (ideal gas), the GENERIC building blocks are the Boltzmann entropy 
\begin{equation}\label{eqn:Boltzmann}
    \uS(f) = -k_B \int \text{d}\rr \text{d}\pp f(\rr,\pp) \left(\log\left(h^3 f(\rr,\pp)\right) - 1\right),
\end{equation}
where $h$ is the Planck constant,
the energy 
\begin{equation}
    \uE(f) = \int \text{d}\rr \text{d}\pp \frac{\pp^2}{2m} f(\rr,\pp) + \frac{1}{2}\int \text{d}\rr \text{d}\rr' \text{d}\pp \text{d}\pp' U(|\rr-\rr'|) f(\rr,\pp) f(\rr',\pp')
\end{equation}
with a pairwise interaction potential $U(|\rr-\rr'|)$,
and the Poisson bracket of the Vlasov equation \cite{grcontmath},
\begin{equation}
    ^\uparrow\{A,B\} = \int \text{d}\rr \text{d}\pp f(\rr,\pp) \left(\frac{\partial A_f}{\partial \rr} \cdot \frac{\partial B_f}{\partial \pp} - \frac{\partial B_f}{\partial \rr} \cdot \frac{\partial A_f}{\partial \pp}\right),
\end{equation}
see for instance \cite{pkg} for more details.
There is no dissipation potential at this level; the detailed dynamics is purely Hamiltonian,

\subsection{Reduction to mass density: The ideal-gas case}\label{sec.diff.ideal}
\paragraph{Reduction mapping.} 
We now assume that there is no interaction potential, $U=0$, so that the system is an ideal gas. 
The goal is to reduce the description to the level of mechanical equilibrium where only the mass and energy densities,
\begin{subequations}
    \begin{align}
    \rho(\rr) =& m \int \text{d}\pp f(\rr,\pp),\\
    e(\rr) =& \int \text{d}\pp \frac{\pp^2}{2m} f(\rr,\pp),
\end{align}
\end{subequations}
are the state variables. Eventually, we shall even assume isothermality, so that only the mass density remains as state variable for simplicity, but let us first keep the energy density field as well. 

\paragraph{MaxEnt and reduced entropy.}
MaxEnt from the Boltzmann entropy \eqref{eqn:Boltzmann} with constraints on $\rho$ and $e$ gives the local Maxwellian distribution
\begin{align}
    f(\rr,\pp) 
    =\frac{1}{h^3}\exp\left(-\frac{1}{k_B}\left(\frac{\pp^2}{2m}e^* + m\rho^*\right)\right)
    = \frac{\rho(\rr)}{m} \left(\frac{e^*}{2\pi m k_B}\right)^{3/2}\exp\left(-\frac{e^*}{2 m k_B} \pp^2\right),
\end{align}
with the inverse temperature $e^* = \frac{3}{2}\frac{k_B}{m e/\rho} = T^{-1}$,
and the corresponding local equilibrium entropy,
\begin{equation}
\dS(\rho,e) = k_B \int \text{d}\rr \frac{\rho(\rr)}{m} \left(\frac{5}{2} + \frac{3}{2}\ln\left(\frac{4\pi m^2}{3h^2}\frac{e}{\rho}\right)-\ln \frac{\rho}{m}\right),
\end{equation}
which is the well known Sackur--Tetrode formula \cite{callen,pkg}.

\paragraph{Hamilton--Jacobi equation.}
The Boltzmann Poisson bracket, when evaluated on functionals of $\rho$ and $e$ only, disappears; there is no mechanical coupling between the mass and energy densities. The Hamilton--Jacobi equation \eqref{eq.HJ.explicit} for the dissipation potential $\Sigma_e^*(\rho^*, e^*)$ thus reduces to
\begin{multline}\label{eq.HJ.ideal}
    \int d\rr \begin{pmatrix}\frac{\delta \Sigma^*_e}{\delta \rho^*} & \frac{\delta \Sigma^*_e}{\delta e^*}\end{pmatrix}
    \begin{pmatrix}
        \frac{5}{2}\frac{k_B}{m \rho} & -\frac{3}{2}\frac{k_B}{m e} \\
        -\frac{3}{2}\frac{k_B}{m e} & \frac{3}{2}\frac{k_B \rho}{m e^2}
    \end{pmatrix}
    \begin{pmatrix}\frac{\delta \Sigma^*_e}{\delta \rho^*} \\ \frac{\delta \Sigma^*_e}{\delta e^*}\end{pmatrix}=\\
    = 
    \int d\rr \frac{\pi^{3/2}\sqrt{m}k_B^{3/2}}{\sqrt{2}h^3(e^*)^{9/2}}\left(35 k_B^2 (\nabla e^*)^2 + 20 k_B m e^* \nabla e^* \cdot \nabla \rho^* + 4m^2 (e^*)^2 (\nabla\rho^*)^2\right)
\end{multline}
where the density and energy density are expressed in terms of the conjugate variables $\rho^*$ and $e^*$,
\begin{equation}
    e = \frac{3}{2}\frac{k_B \rho}{m e^*} 
    \quad\text{and}\quad
    \rho = m \exp\left(-\frac{m \rho^*}{k_B} \right)\left(\frac{2\pi m k_B}{h^3}\right)^{3/2} \frac{1}{(e^*)^{3/2}}.
\end{equation}
Linearizing around a homogeneous reference state $\rho\approx \rho_0$, and assuming isothermality ($e^*\approx 1/T$, $T$ being constant), the Hamilton--Jacobi equation \eqref{eq.HJ.ideal} for $\Sigma_e^*$ becomes
\begin{equation}\label{eq.diff.hj.id}
    \int d\rr \frac{2}{5}\frac{m \rho_0^2 T}{k_B }(\nabla \rho^*)^2 = \int d\rr \left(\frac{\delta \Sigma_e^*}{\delta \rho^*}\right)^2.
\end{equation}

\paragraph{Operator solution.} 
We seek a linear solution $\frac{\delta \Sigma_e^*}{\delta \rho^*} = K \rho^*$ so that the dissipation potential is convex ($K$ is thus an unknown linear operator). Substituting into \eqref{eq.diff.hj.id} gives the operator equation
\begin{equation}
    K^2 = \frac{2}{5}\frac{m \rho_0^2 T}{k_B }(-\Delta),
\end{equation}
with $\Delta$ the Laplace operator, giving
\begin{equation}\label{eq.K.ideal}
    K = \sqrt{\frac{2 m \rho_0^2 T}{5 k_B}} \sqrt{-\Delta},
\end{equation}
see Appendix \ref{appen:spectral} for details on the square root of negative Laplacian.
The dissipation potential and the resulting evolution equation are
\begin{equation}
    \Sigma_e^*(\rho^*) = \frac{1}{2}\sqrt{\frac{2 m \rho_0^2 T}{5 k_B}} \int d\rr \rho^* \sqrt{-\Delta}\, \rho^*,
    \qquad
    \partial_t \rho = K\rho^* = \sqrt{\frac{2 m \rho_0^2 T}{5 k_B}}\, \sqrt{-\Delta}\, \rho^*.
\end{equation}
However, the operator $\sqrt{-\Delta}$ is \emph{nonlocal}: in Fourier space its symbol is $2\pi|\xxi|$, which cannot be written as a differential operator of finite order. Therefore, although a well-defined dissipation potential exists, the reduced evolution is not a diffusion equation. This is the key negative result for the ideal gas, which we shall overcome by adding interactions and thus introducing an interaction length scale into the system, as we show in the next section.

\paragraph{Numerical signature of nonlocal diffusion.}
In a particle simulation one would observe this nonlocality as an anomalous, non-Gaussian spreading of the density profile. For example, imagine a localized density perturbation in a large periodic box and compute $\rho(\rr,t)$ from particle histograms. Classical diffusion predicts Gaussian profiles and mode decay rates proportional to $|\xxi|^2$, whereas the operator $\sqrt{-\Delta}$ generates heavier spatial tails and a distinct scaling of low-wavenumber modes. Equivalently, in Fourier space the decay rate of each mode is proportional to $|\xxi|$ (for the linearized relation $\rho^*\propto \rho$), not $|\xxi|^2$, so a log-linear plot of $\log |\widehat{\rho}(\xxi,t)|$ versus time at fixed $\xxi$ would reveal the $|\xxi|$ scaling of the relaxation rate. Detecting this non-quadratic dispersion in particle simulations would thus be a direct numerical signature of the nonlocal diffusion predicted by the lack-of-fit reduction. It is, however, out of scope of the present paper, and we leave it for future work.

\subsection{Reduction with interactions: Recovery of diffusion}\label{sec.diff.interaction}
The physical reason for the nonlocality above is that non-interacting particles have no intrinsic length scale that could localize the momentum transfer. We now show that introducing short-range interactions provides such a scale and makes the operator local.

\paragraph{Cahn--Hilliard free energy.}
We add an interparticle potential $U(|\rr-\rr'|)$ of hard-core type ($U=\infty$ for $r<d$, $U\leq k_B T$ for $r>d$). The free energy acquires a gradient term of Cahn--Hilliard type \cite{pismen},
\begin{equation}\label{eq.F.CH}
    \dF(\rho,T) = \int d\rr\, f_0(\rho, T) + \int d\rr \frac{1}{2m^2} k |\nabla \rho|^2,
    \qquad
    k = -\frac{2 \pi k_B T}{3} \int_d^\infty dr\, r^4 U(r),
\end{equation}
where $f_0(\rho,T)$ is the bulk free energy density (present when $\rho$ becomes homogenous). For instance, the van der Waals gas would have 
\begin{equation}
    f_0(\rho,T) = f_{id}(\rho,T)
    +k_B T \left(b-\frac{a}{T} \right)\int d\rr \frac{\rho^2}{2 m^2}
    \quad\text{with}\quad
    f_{id}=k_B T \frac{\rho}{m}\left(\ln\left(\frac{\rho}{m}\right) + \ln\Lambda^3 -1 \right)
\end{equation} 
and $\Lambda = h/\sqrt{2\pi m k_B T}$, see \cite{callen,landau5,pismen}. 

Our goal will be to obtain the lower conjugate entropy $\dS^*$ and take the inverse of its second differential as the Hessian of the lower entropy $\dS$ itself. The lower conjugate entropy $\dS^*(\yy^*)$ is tighly connected with the free energy be Legendre transformation, 
\begin{equation}
    \dS^*(\yy^*) = \frac{\Omega}{T} = \frac{\dF - \mu \int d\rr\, \rho}{T},
\end{equation}
where $\Omega$ is the grand potential and $\mu$ the chemical potential \cite{landau5}. To carry out the Legendre transformation from $\rho$ to $\rho^*$, we approximate the free energy to quadratic degree in $(\rho-\rho_0)/\rho_0$ around a homogeneous reference state $\rho_0$. Free energy \eqref{eq.F.CH} in the quadratic approximation becomes
\begin{equation}
    \dF(\rho,T) \approx \alpha + \int d\rr \beta \rho + \frac{1}{2}\gamma \int d\rr \rho^2  + \frac{1}{2}\int d\rr \frac{1}{m^2} k |\nabla \rho|^2,
\end{equation}
where the coefficients are
\begin{subequations}
\begin{align}
    \alpha(\rho_0,T) =& \int d\rr f_{id}(\rho_0,T) - k_BTV\frac{\rho_0}{m} \ln\Lambda^3 + \frac{1}{2}k_B T V \frac{\rho_0}{m}\\
    \beta(\rho_0,T) =& \frac{k_B T}{m} \left(\ln\Lambda^3 -1\right)\\
    \gamma(\rho_0,T) =& \frac{k_B T}{m \rho_0} + \frac{k_B T}{m^2} \left(b-\frac{a}{T} \right).
\end{align}
\end{subequations}

\paragraph{Conjugate variables and the operator $L$.}
The Legendre transformation gives the relation between $\rho$ and its conjugate $\rho^*$:
\begin{equation}\label{eq.rhosrho}
    \rho = L^{-1}\left(\frac{-\rho^*T-\beta}{\gamma}\right), \qquad \rho^* = -\frac{\gamma}{T} L \rho - \frac{\beta}{T},
\end{equation}
where $L = 1 - l^2 \nabla^2$ is a self-adjoint positive-definite operator with diffusion length $l = \sqrt{k/\gamma}/m$. The conjugate entropy becomes
\begin{equation}
    \dS^*(\rho^*, T) = \frac{\alpha}{T}V - \frac{\gamma}{2T} \int d\rr \left(\frac{-\rho^*T-\beta}{\gamma}\right)L^{-1}\left(\frac{-\rho^*T-\beta}{\gamma}
    \right).
\end{equation}
The second functional derivative of $\dS^*$ with respect to $\rho^*$ is thus given by 
\begin{equation}
    \frac{\delta^2 \dS^*}{\delta \rho^* \delta \rho^*} = -\frac{T}{\gamma} L^{-1},
\end{equation}
and the inverse operator is then the Hessian of the lower entropy $\dS$, 
\begin{equation}
    \frac{\delta^2 \dS}{\delta \rho \delta \rho} = -\frac{\gamma}{T} L.
\end{equation}  

\paragraph{Modified Hamilton--Jacobi equation.}
Now, the right-hand side of the Hamilton--Jacobi equation \eqref{eq.diff.hj.id} changes, as the identity operator is replaced by $L$, giving
\begin{equation}\label{eq.hj.interaction}
    \int d\rr \frac{2}{5}\frac{m \rho_0^2 T}{k_B }(\nabla \rho^*)^2 = \int d\rr \left(\frac{\delta \Sigma_e^*}{\delta \rho^*}\right) L \left(\frac{\delta \Sigma_e^*}{\delta \rho^*}\right).
\end{equation}
Note that for simplicity we neglect the possible changes on the left hand side due to the presence of the potential in the Vlasov equation.
With the linear ansatz $\frac{\delta \Sigma_e^*}{\delta \rho^*} = K \rho^*$, the operator equation becomes
\begin{equation}
    K L K = \frac{2}{5}\frac{m \rho_0^2 T}{k_B }(-\Delta).
\end{equation}
Assuming that $K$ and $L$ commute, the solution is
\begin{equation}\label{eq.K.interaction}
    K = \sqrt{\frac{2m \rho_0^2 T}{5 k_B}}\, L^{-1/2} \sqrt{-\Delta},
\end{equation}
see Appendix \ref{appen:spectral} for details.
Compared to the ideal-gas result \eqref{eq.K.ideal}, the factor $L^{-1/2}$ is the crucial new ingredient.

\paragraph{Localization at short wavelengths.}
In Fourier space, the symbols of the two operators are $\widehat{\sqrt{-\Delta}} = 2\pi |\xxi|$ and $\widehat{L^{1/2}} = \sqrt{1+l^2 4\pi^2 |\xxi|^2}$. For wavelengths shorter than the interaction range ($l\,|\xxi| \gg 1$), the latter simplifies to $\widehat{L^{1/2}} \approx 2\pi l |\xxi|$.

The evolution equation for $\rho$ is then given by 
\begin{equation}
    \partial_t \rho = \sqrt{\frac{2m \rho_0^2 T}{5 k_B}} L^{-1/2} \sqrt{-\Delta}\, \rho^*
    \stackrel{\text{\eqref{eq.rhosrho}}}{=} -\frac{\gamma}{T}\sqrt{\frac{2m \rho_0^2 T}{5 k_B}} 
    L^{-1/2} \sqrt{-\Delta}\, L \rho.
\end{equation}
In the short-wavelength limit, where
\begin{equation}
    \widehat{L^{-1/2} \sqrt{-\Delta} L} = \widehat{L^{1/2} \sqrt{-\Delta}} \approx 2\pi l |\xxi| \cdot 2\pi |\xxi| = 4\pi^2 l |\xxi|^2 = l\widehat{-\Delta},
\end{equation}
this reduces to the classical diffusion equation
\begin{equation}
    \partial_t \rho =  D \Delta \rho,
\end{equation}
with the diffusion coefficient
\begin{equation}
    D = \sqrt{\frac{2m \rho_0^2}{5 k_B T}}\,\gamma\, l.
\end{equation}  
When we use the relations for $\gamma$ and $l$ in terms of the parameters of the free energy, we obtain the diffusion coefficient
\begin{equation}
    D = \sqrt{\frac{2k}{5m}} \sqrt{\frac{\rho_0}{m}} \sqrt{1 + \frac{\rho_0}{m}\left(b-\frac{a}{T} \right)},
\end{equation}
It should be noted, however, that this diffusion coefficient is not the be interpreted as the self-diffusion coefficient of a tagged particle \cite{landau10}, but rather as a coefficient giving the overdamped evolution of mass density when the momentum degrees of freedom are not resolved and have relaxed to a local equilibrium. 

\paragraph{Summary.} The lack-of-fit reduction of the Vlasov equation to mass density always yields a well-defined dissipation potential. For an ideal gas, the corresponding operator is nonlocal and the evolution is not diffusive. When short-range interactions are present, they introduce a length scale $l$ that localizes the operator, and the classical diffusion equation emerges. This provides an independent derivation of diffusive transport from a purely Hamiltonian kinetic equation, complementing the finite-dimensional Kac--Zwanzig example of the preceding section.

\section{Conclusion}
In this paper, we have shown how the arrow of time emerges from the combination of two conditions: an ergodic or phase-mixing detailed system, which stays close to the MaxEnt submanifold, and incomplete knowledge of its precise state. When the only a reduced level of description is retained and the original Hamiltonian system stays close to the MaxEnt submanifold corresponding to the reduced description, the lack-of-fit reduction method reveals the effective irreversible evolution equations for the reduced variables. To establish this rigorously, the lack-of-fit reduction is reformulated in the context of the Onsager-Machlup principle, minimizing the information discrepancy between the detailed and reduced evolution vector fields.

To write the Onsager-Machlup action, we introduced a \emph{MaxEnt Kullback--Leibler discrepancy} that measures the information distance between two manifolds related by the principle of maximum entropy (MaxEnt).
The MaxEnt discrepancy generalizes the concept of Kullback–Leibler divergence and Fisher information matrix to arbitrary concave entropies. In the case of affine reduction mappings, the MaxEnt discrepancy reduces to the Bregman distance, but it applies also to nonlinear reducdtion mappings.

The resulting lack-of-fit reduction method then leads to reduced dynamics of the GENERIC form, which consists of projected Hamiltonian mechanics and gradient dynamics with an \emph{emergent dissipation potential} which encodes the discrepancy between the detailed (unresolved) evolution and the reduced evolution equations. 

The method is then tested on the Kac–Zwanzig model, showing a qualitative agreement with the numerical results (with no fitting parameters). In particular, even purely Hamiltonian dynamics can be reduced to a system with fewer degrees of freedom, and dissipation then emerges on the reduced level of description, without any fitting parameters. Lack of fit reduction thus provides a systematic way to derive effectively dissipative evolution equations of reduced systems even from purely Hamiltonian detailed dynamics.

Three levels of the lack-of-fit reduction are calculated in the context of the Kac–Zwanzig model, differing in the choice of the reduced state variables. The best agreement between the reduced and detailed dynamics is obtained for the set of reduced state variables that has the lowest KL divergence between the distributions of the detailed and reduced dynamics. Another clue for how to choose the reduced state variables is how much the resulting reduced dynamics corrects the projected Hamiltonian dynamics. The less correction is needed, the better the choice of the reduced state variables, and in the Kac--Zwanzig case, the choice with lower KL divergence has also lower reversible correction in the dissipative dynamics, see Appendix \ref{appen:detailbalance}.

A path-integral formulation is provided, following \cite{kleeman}, which turns the reduced dynamics to a Langevin equation compatible with the lack-of-fit action. This stochastic extension captures some fluctuations of the reduced dynamics caused by the unresolved degrees of freedom.

Finally, the lack-of-fit reduction method is illustrated on the example of diffusion emerging from the Vlasov equation when interactions between particles are present. It is shown that the dissipation potential exists in both cases, with and without interactions, but it can be approximated by a local operator (diffusion) when interactions are present.

In the future, we would like to focus on problems with boundary conditions, on a large-deviation generalization of the method, and on applications to systems in continuum thermodynamics, including viscosity.

\section*{Acknowledgements}
The authors were supported by the Czech Science Foundation (GACR) under project 23-05736S. MP is a member of the Nečas Center for Mathematical Modeling. KM is supported by Charles University, project GA UK No. 414225.
MP is grateful to Celia Reina, Mark Peletier, Vojtěch Votruba, Jan Engler, Martin Šípka, and Šimon Andrš for discussions on the topic of this paper. MP acknowledges the use of Claude Opus 4.6 for cleaning the manuscript and for speeding up the typesetting.

\appendix

\section{Details on the MaxEnt calculation}\label{sec.MaxEnt.detail}
Let us recall a few rather technical but useful relations \cite{JSP2020}. Taking another derivative of the MaxEnt equation \eqref{eqn:duS}, we obtain that
\begin{equation}\label{eq.d2uS}
\frac{\partial^2 \uS}{\partial x^i \partial x^j}\Big|_{\xx(\yy^*)} \frac{\partial x^j}{\partial y^*_b} 
= \frac{\partial \pi^b}{\partial x^i}\Big|_{\xx(\yy^*)} + \frac{\partial^2 \pi^a}{\partial x^i \partial x^j}\Big|_{\xx(\yy^*)} \frac{\partial x^j}{\partial y^*_b} y^*_a.
\end{equation}
In the case of affine mapping $\pi$, the second term vanishes. 

Another useful formula shows how the Hessian of the reduced conjugate entropy relates to the Hessian of the detailed entropy. The first derivative of the definition of the reduced conjugate entropy \eqref{eqn:dSstar} gives
\begin{equation}
   \frac{\partial \dS^*}{\partial y^*_a} 
   = -\frac{\partial \uS}{\partial x^i}\frac{\partial x^i}{\partial y^*_a} + \pi^a
   + \frac{\partial \pi^b}{\partial x^i}\Big|_{\xx(\yy^*)} \frac{\partial x^i}{\partial y^*_a} y^*_b.
\end{equation}
The second derivative is then
\begin{align}\label{eqn.d2Sstar2}
    \frac{\partial^2 \dS^*}{\partial y^*_a \partial y^*_b} =& 
    -\frac{\partial^2 \uS}{\partial x^i \partial x^j}\Big|_{\xx(\yy^*)} \frac{\partial x^i}{\partial y^*_a} \frac{\partial x^j}{\partial y^*_b} 
    -\frac{\partial \uS}{\partial x^i}\Big|_{\xx(\yy^*)} \frac{\partial^2 x^i}{\partial y^*_a\partial y^*_b} 
    + \frac{\partial \pi^a}{\partial x^i }\Big|_{\xx(\yy^*)} \frac{\partial x^i}{\partial y^*_b}
    \nonumber\\
    &+ \frac{\partial^2 \pi^c}{\partial x^i \partial x^j}\Big|_{\xx(\yy^*)} \frac{\partial x^i}{\partial y^*_a} \frac{\partial x^j}{\partial y^*_b} y^*_c 
    + \frac{\partial \pi^c}{\partial x^i }\Big|_{\xx(\yy^*)} \frac{\partial^2 x^i}{\partial y^*_a\partial y^*_b}  y^*_c 
    + \frac{\partial \pi^b}{\partial x^i }\Big|_{\xx(\yy^*)} \frac{\partial x^i}{\partial y^*_a}.
\end{align}
In order to simplify the right-hand side of this expression, we take derivative of the equality
\begin{equation}
    \pi(\xx(\yy^*(\yy))) = \yy,
\end{equation}
which tells us that the projection $\pi$ of the MaxEnt mapping is the identity. After the differentiation, we obtain
\begin{equation}
    \frac{\partial \pi^a}{\partial x^i}\Big|_{\xx(\yy^*(\yy))} \frac{\partial x^i}{\partial y^*_a}\frac{\partial y^*_a}{\partial y^b} = \delta^a_b, 
\end{equation}
which means that 
\begin{equation}
    \frac{\partial \pi^a}{\partial x^i}\Big|_{\xx(\yy^*)} \frac{\partial x^i}{\partial y^*_b} = \frac{\partial y^a}{\partial y^*_b} = \frac{\partial^2 \dS^*}{\partial y^*_a \partial y^*_b}.
\end{equation}
Equality \eqref{eqn.d2Sstar2} can be simplified to
\begin{equation}\label{eq.gab.d2S}
   -g^{ab} = \frac{\partial^2 \dS^*}{\partial y^*_a \partial y^*_b} + y^*_c \frac{\partial^2 \pi^c}{\partial x^i \partial x^j}\Big|_{\xx(\yy^*)} \frac{\partial x^i}{\partial y^*_a} \frac{\partial x^j}{\partial y^*_b},
\end{equation}
where $g^{ab}$ is the Fisher information matrix \eqref{eq.gab}.

\section{MaxEnt KL discrepancy with Tsallis-Havrda-Charvát entropy}\label{sec.tsallis}
In this section, we show how the generalized MaxEnt KL discrepancy \eqref{eqn:DKLM} works in the case of the Tsallis-Havrda-Charvát entropy \cite{tsallis,havrda-charvat}, 
\begin{equation}
    \uS_q^{\mathrm{Tsallis}}(\pp) = -\sum_i p_i \log_q p_i = -\sum_i p_i \frac{p_i^{1-q}-1}{1-q},
\end{equation}
where $q$ is the entropic index, parametrizing a family of entropies, and $\log_q(x) = \frac{x^{1-q}-1}{1-q}$. This entropy is a powerful generalization of the Boltzmann-Gibbs-Shannon entropy (recovered for $q=1$), as its MaxEnt distributions fit a large class of physical phenomena \cite{tsallis-book}.

We assume that the detailed variables are given by a distribution with probabilities $p_i$ while the reduced variables are given by probabilities $q_i$. 
First, we have to calculate the reducing potential,
\begin{equation}
 \dSt^*(\pp,\qq^*) = \frac{1}{1-q} \sum_i p_i \left(p_i^{1-q}-1\right) + \sum_i q^*_i p_i.
\end{equation}
The potential has a minimum at $p_i(\qq^*) = \left(\frac{1-(1-q)q^*_i}{2-q}\right)^{1/(1-q)}$, and when evaluated at this minimum, it gives the lower conjugate entropy
\begin{equation}
    \dS^*(\qq^*) = \sum_i \left(\frac{(q-1)q^*_i + 1}{2-q}\right)^{\frac{2-q}{1-q}}.
\end{equation}
The consequent Legendre transformation 
    $q_i = \frac{\partial \dS^*}{\partial q^*_i}$
gives a relation $q^*_i = \frac{(2-q)q_i^{1-q}-1}{q-1}$.

The MaxEnt KL discrepancy, which here becomes a divergence, then becomes
\begin{equation}\label{eq.DKL.Tsallis}
    D^M_KL(\pp||\qq) = \sum_i p_i \left(\log_q p_i - \log_q q_i\right) 
    + (1-q)\sum_i (q_i-p_i) \log_q q_i.
\end{equation}
In the case when $q=1$, the KL discrepancy simplifies to the standard KL divergence for Shannon entropy. Formula \eqref{eq.DKL.Tsallis} is, however, different from the usual KL divergence for the Tsallis entropy \cite{furuichi},
\begin{equation}
    D^M_{KL}(\pp||\qq) \neq -\sum_i p_i \log_q\frac{p_i}{q_i}.
\end{equation}
For $q=1$, the two divergences coincide and simplify to the standard KL divergence for Shannon entropy.
As the generalized KL discrepancy is only a by-product of this manuscript, we leave further analysis for future work.

\section{Asymptotic solution of the Riccati equation}\label{appen:riccati}
In this appendix, we provide the full asymptotic analysis of the stationary Riccati equation \eqref{eqn:Riccati} for the Kac--Zwanzig model. We rescale the matrices into the following non-dimensional form (omitting the common prefactor):
\begin{equation}
\tilde{\mathbb{A}}
=
\begin{pmatrix}
1 & 0 & -1\\
0 & 1+b &0  \\
-1  & 0  & 1+\rho
 \end{pmatrix}, \hskip0.3cm
 \tilde{\mathbb{B}}
=
\begin{pmatrix}
0 & 1 & 0\\
-1 & 0 & 1 \\
0 & -1 & 0 
 \end{pmatrix}, \hskip0.3cm
\tilde{\mathbb{C}}
=
\begin{pmatrix}
b^{-1} & 0 & 0\\
0 & 1 & 0\\
0 & 0 & 1
 \end{pmatrix},
\end{equation}
where $\rho=b \frac{\overline{\omega^2}}{\alpha/M}$ with $0<b=\frac{\alpha}{\gamma}$. We have omitted the fourth row and column for simplicity. For asymptotic calculations, we consider $1\ll\rho$ while $b=\mathcal{O}(1)$.

The algebraic Riccati equation becomes the following set of quadratic equations for $\tilde{\mathbb{M}}$:
\begin{subequations}
\begin{align} \label{eq.MalgSetApp}
    0&=-\tilde{m}_{12}(2+\tilde{m}_{12})-\tilde{m}_{13}^2-\tilde{m}_{11}^2 b,\\
    0&=-(1+\tilde{m}_{12})\tilde{m}_{22}-\tilde{m}_{13}(1+\tilde{m}_{23})+\tilde{m}_{11} (1-\tilde{m}_{12}) b,\\
    0&=\tilde{m}_{12}-\tilde{m}_{23}(1+\tilde{m}_{12})-\tilde{m}_{13}(\tilde{m}_{11} b+\tilde{m}_{33}),\\
    0&=-\tilde{m}_{22}^2-\tilde{m}_{23}(2+\tilde{m}_{23})+(2-\tilde{m}_{12}) \tilde{m}_{12} b,\\
    0&=\tilde{m}_{22}(1-\tilde{m}_{23})-(1+\tilde{m}_{23})\tilde{m}_{33}+(1-\tilde{m}_{12})\tilde{m}_{13} b,\\
    0&=(2-\tilde{m}_{23})\tilde{m}_{23}-\tilde{m}_{33}^2-\tilde{m}_{13}^2 b + \rho,
\end{align}
\end{subequations}
where the symmetry of the dissipative potential has been employed. The dimensional dissipation matrix follows from
\begin{equation*}
    \mathbb{M}= \begin{pmatrix}
\rho~\tilde{m}_{11} & \tilde{m}_{12} & \rho~\tilde{m}_{13}\\
\tilde{m}_{12} & \rho^{-1} \tilde{m}_{22} & \tilde{m}_{23}\\
\rho~\tilde{m}_{13} & \tilde{m}_{23} & \rho~\tilde{m}_{33}
 \end{pmatrix}.
\end{equation*}

To find an asymptotic solution for $\rho\gg 1$, the correct asymptotic orders are determined by dominant balance:
\begin{align*}
\tilde{m}_{11} &= \mathcal{O}(1), \tilde{m}_{12} = \mathcal{O}(1), \tilde{m}_{13} = \mathcal{O}(\rho^{-1/2}), \\
\tilde{m}_{22} &= \mathcal{O}(1), \tilde{m}_{23} = \mathcal{O}(1), \tilde{m}_{33} = \mathcal{O}(\rho^{1/2}). 
\end{align*}
This yields eight leading-order solutions for the dimensional problem:
\begin{align*}
    m_{11} &= (M\alpha)^{-1/2} z_1 \left(\left(\frac{1}{b}-4-8b\right)+8z_2\sqrt{b(1+b)}\right)^{1/2},\\
    m_{12} &= -1-2b+2z_2\sqrt{b(1+b)},\\
    m_{13} &= (M\alpha)^{-1/2} z_3 \left(z_2(1+4b)-b\sqrt{b(1+b)}\right)^{1/2}\rho^{-1/2},\\
    m_{22} &= (M\alpha)^{1/2} z_1 z_2 \sqrt{b(1+b)} \left(\left(\frac{1}{b}-4-8b\right)+8z_2 \sqrt{b(1+b)}\right)^{1/2},\\
    m_{23} &= -1,\\
    m_{33} &= -(M\alpha)^{-1/2} z_2 z_3 \rho^{1/2},
\end{align*}
where $z_j\in\{-1,1\}$ for all $j$, giving $2^3=8$ solutions. Positive semidefiniteness of the dissipation matrix requires $z_1=+1=z_2$, $z_3=-1$, yielding the unique result \eqref{eq.Masymptotic}.

\paragraph{Numerical comparison.} For $M=1099$, $\alpha=1$, $b=1$, $\rho=37$, the asymptotic estimate gives
\begin{equation*}
    \mathbb{M}_{\text{asymp}} = \begin{pmatrix}
        0.0168952 & -0.171573 & 0.0198139 \\ & 26.2589 & -1& \\ & & 0.0301648
    \end{pmatrix},
\end{equation*}
while the full numerical solution of the Riccati equation reads
\begin{equation*}
    \mathbb{M}_{\text{num}} = \begin{pmatrix}
        0.0166733 & -0.1696847 & 0.002148 \\ & 25.1330519 & -0.76117& \\ & & 0.179169
    \end{pmatrix}.
\end{equation*}
The asymptotic estimate is in good agreement with the numerical solution.

\section{Convexity of the effective dissipation potential}\label{appen:convexity}
In this section, we show how the convexity of the effective dissipation potential $\Psi_e=-\Sigma_e^*$ follows from the convexity of the cotangent-bundle Lagrangian $\mathcal{L}^*$. 

Fix an initial time $t$ and a terminal time $T$, and consider two minimizing paths $\yy^{*1}(s)$ and $\yy^{*2}(s)$ on the interval $[t,T]$ that have different initial points $\yy^{*1}(t)$ and $\yy^{*2}(t)$ but the same terminal point $\yy^{*1}(T)=\yy^{*2}(T)=0$. Define the interpolated path
\begin{equation}
    \yy^{*\lambda}(s) = (1-\lambda)\yy^{*1}(s) + \lambda \yy^{*2}(s),
    \qquad \lambda\in[0,1],
\end{equation}
which has initial point $\yy^{*\lambda}(t) = (1-\lambda)\yy^{*1}(t) + \lambda \yy^{*2}(t)$ and the same terminal point $\yy^{*\lambda}(T)=0$.

The action along the interpolated path is given by
\begin{equation}
    \int_t^T \mathcal{L}^*(\yy^{*\lambda}(s), \dot{\yy}^{*\lambda}(s)) \, ds.
\end{equation}
From the convexity of the Lagrangian, expressed by Jensen's inequality, we have
\begin{equation}
    \int_t^T \mathcal{L}^*(\yy^{*\lambda}(s), \dot{\yy}^{*\lambda}(s)) \, ds
    \leq (1-\lambda)\Psi_e(t,\yy^{*1}(t)) + \lambda \Psi_e(t,\yy^{*2}(t)),
\end{equation}
where $\Psi_e$ on the right-hand side is the action evaluated along the minimizing paths $\yy^{*1}(s)$ and $\yy^{*2}(s)$, respectively. Taking the minimum of the left-hand side over all paths with the same initial point $\yy^{*\lambda}(t)$ and terminal point $0$, we obtain
\begin{equation}
    \Psi_e\bigl(t,(1-\lambda)\yy^{*1}(t)+\lambda\yy^{*2}(t)\bigr)
    \leq (1-\lambda)\Psi_e(t,\yy^{*1}(t)) + \lambda \Psi_e(t,\yy^{*2}(t)),
\end{equation}
which means that the effective dissipation potential $\Psi_e$ is convex in the initial cotangent state.

\section{Details on detailed balance}\label{appen:detailbalance}

In this section, we give details on the time-antisymmetric part of the lack-of-fit Lagrangian. For time-reversed trajectories it is sufficient to track the transformation of $\dot{\yy}-\YY(\Sigma_e,\yy)$ and $\gmetric$. The parity under time reversal is denoted $p(f)$. We decompose
\begin{equation}
    \Sigma_e = \frac{1}{2}\left(\Sigma_e^{\textrm{even}} + \Sigma_e^{\textrm{odd}}\right),
    \qquad
    \Theta(\Sigma_e) = \frac{1}{2}\left(\Sigma_e^{\textrm{even}} - \Sigma_e^{\textrm{odd}}\right).
\end{equation}
Then
\begin{equation}
    \Theta \left[g^{ab}\frac{\partial \Sigma_e}{\partial y^b} \right]
    = p(y^a)\left(g^{ab}\frac{\partial \Sigma^{\textrm{even}}_e}{\partial y^b} - g^{ab}\frac{\partial \Sigma^{\textrm{odd}}_e}{\partial y^b}\right).
\end{equation}
We assume that the even part is annihilated by the resolved Poisson bracket, i.e.
\begin{equation}
    \frac{\partial \Sigma^{\textrm{even}}_e}{\partial y^a}\, \dL^{ab} \frac{\partial \dE}{\partial y^b} = 0.
\end{equation}
With this, the zero-cost trajectory can be split into reversible and irreversible parts:
\begin{equation}
    \Theta \left[Y^a(\Sigma_e, \yy ) \right]
    = - p(y^a) \left(\dL^{ab} \frac{\partial \dE}{\partial y^b} + g^{ab}\frac{\partial \Sigma^{\textrm{odd}}_e}{\partial y^b} \right)
    + p(y^a) \left(\underbrace{\frac{\partial \pi^a}{\partial x^i}\Big|_{\xx(\yy^*)}\frac{\partial ^{\uparrow}  \Xi}{\partial x^*_i} \Bigg|_{\mathbf{x}^* = \frac{\partial \uS}{\partial {\mathbf{x}}}(\xx(\yy^*))}}_{\Delta_{\textrm{irr}} \pi^a} + g^{ab}\frac{\partial \Sigma^{\textrm{even}}_e}{\partial y^b} \right).
\end{equation}
The expression $\Delta_{\textrm{irr}} \pi^a$ indicates the irreversible change of the reduced variable caused by the detailed dissipation $^{\uparrow}\Xi$; note that $\pi^a(\xx(\yy^*(\yy))) = y^a$.

Returning to the full Lagrangian, the time-antisymmetric part is
\begin{equation}
    \mathcal{L}(\Theta \yy, \Theta \dot{\yy}) - \mathcal{L}(\yy, \dot{\yy})
    = 2 \underbrace{\left(\dot{y}^a - \dL^{ab} \frac{\partial \dE}{\partial y^b} - g^{ab}\frac{\partial \Sigma^{\textrm{odd}}_e}{\partial y^b}\right)}_{\dot{y}_{\textrm{irr}}}
    \left( g_{ab} \Delta_{\textrm{irr}} \pi^b +  \frac{\partial \Sigma^{\textrm{even}}_e}{\partial y^a} \right).
\end{equation}

By $\dot{y}_{\textrm{irr}}$ we mean the irreversible evolution of the reduced variable $\yy$.

\begin{figure}[ht!]
\centering
\includegraphics[scale=0.2]{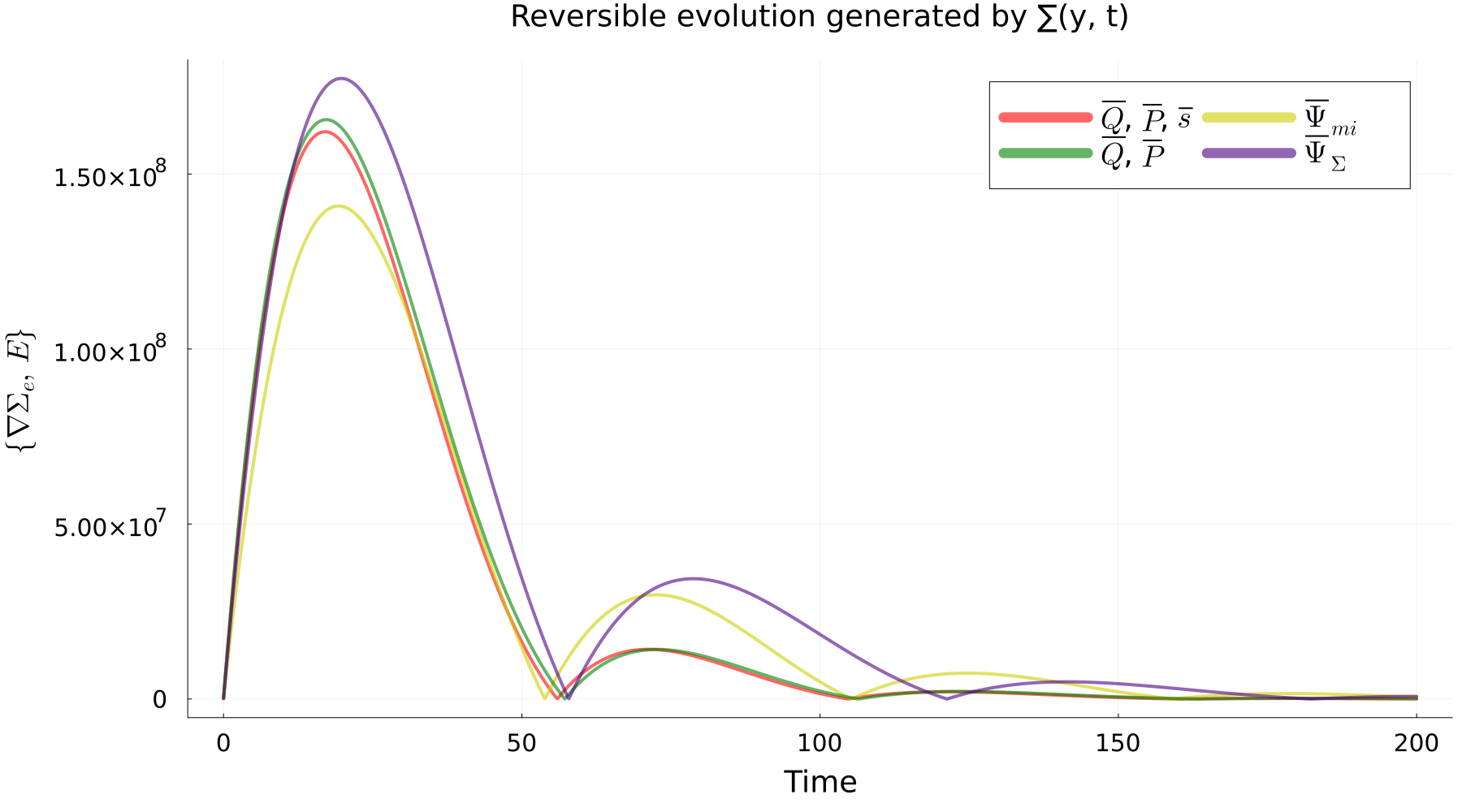}
\caption{Comparison of the reversible evolution generated by the solution of the Hamilton-Jacobi (Riccati) $\Sigma(y,t)$ equation for different choices of the reduced state variables – for two state variables $\QQ$, $\PP$ (green), three with additional $s$ (red) and four variables with the extra variables $\Psi_{mi}$ (yellow) and $\Psi_\Sigma$ (purple) \eqref{eq.extra}. The initial positions of the big particle is $Q_0=1000$ and all other reduced variables initiate at $0$. Parameters of the simulation were $N=10000$, $\gamma=1.0$, $\alpha = 1.0$, $\omega_i$ sampled from uniformly from the interval $(\sqrt{1.0E-05}, \sqrt{1.0E-01})$, and $M$ was set equal to the total mass of the small particles.}
\label{fig.reversibleSigma}
\end{figure}

The dynamics found through the Hamilton-Jacobi equation is thus both reversible (generated by $\Sigma^{\textrm{odd}}_e$) as well as dissipative. To illustrate the relevance of this reversible part in $\Sigma_e$, we plot the evolution of the observable $\frac{\partial \Sigma^{\textrm{odd}}_e}{\partial y^b}$ for the Kac-Zwanzig example in Figure \ref{fig.reversibleSigma}. It is interesting to note that higher values might correlate with unsuitable choice of reduced variables, as viewed by comparing the two choices of the fourth variable ($\Psi_\Sigma$ and $\Psi_{mi}$).

%\textcolor{red}{Compare with the symmetry for detailed evolution $$\mathcal{L}(\Theta \xx, \Theta \dot{\xx}) - \mathcal{L}(\xx, \dot{\xx}) = 2 \dot{\xx}^i_{\textrm{irr}} \frac{\partial \uS }{\partial \xx^i} = 2 \dot{\xx}^i_{\textrm{irr}} \xx^*_i$$}

%\textcolor{red}{By comparison, there should be a function $ \mathcal{S}$ such that $$\mathcal{L}(\Theta \yy, \Theta \dot{\yy}) - \mathcal{L}(\yy, \dot{\yy}) = 2 \dot{y}^a_{\textrm{irr}} \frac{\partial \mathcal{S} }{\partial \yy^a}, \hskip0.2cm \frac{\partial \mathcal{S} }{\partial \yy^a} = \Delta t \frac{\partial}{\partial \yy^a} \left(\frac{\partial \dS }{\partial \yy^b} \Delta_{\textrm{irr}} \pi^b + \Sigma^{\textrm{even}}_e \right).$$}

\section{Calculation of the Kullback–Leibler divergence} \label{append:KLDivCalc}

With the choices of variables given in Table \ref{table:KLdiv}, each result has the form:

$$\log\left(\frac{\mathcal{Z}(\mathbf{z}^*)}{\mathcal{Z}(\mathbf{y}^*)}\right)\frac{1}{k_B} \left( \left< \frac{\delta \mathbf{z}}{\delta f} \right>_{\mathbf{y}} \cdot \mathbf{z}^* -  \left< \frac{\delta \mathbf{y}}{\delta f} \right>_{\mathbf{y}} \cdot \mathbf{y}^* \right),$$ with $\left<  \right>_{\mathbf{y}}$ denoting the mean value with respect to the distribution $\tilde{f}(x;\mathbf{y}^*)$ where $\mathcal{Z}$ is a number dependent on the parameters of the model, e.g. masses of particles etc.,
$$\mathcal{Z}(\mathbf{y}^*) = \mathcal{Z} \exp{\left( - \frac{T \Sigma(\mathbf{y}^*)}{2k_B}\right)} \equiv \mathcal{Z} \exp{\left( \frac{1}{2k_B} \left< \frac{\delta \mathbf{y}}{\delta f} \right>_{\mathbf{y}} \cdot \mathbf{y}^* \right)}.$$ 
This allows us to write
$$D_\text{KL}(\tilde{f}(x;\mathbf{y}^*) \parallel \tilde{f}(x;\mathbf{z}^*)) = \frac{1}{2 k_B^2} \left( \left< \frac{\delta \mathbf{z}}{\delta f} \right>_{\mathbf{y}} \cdot \mathbf{z}^* -  \left< \frac{\delta \mathbf{y}}{\delta f} \right>_{\mathbf{y}} \cdot \mathbf{y}^* \right) \left( \left< \frac{\delta \mathbf{z}}{\delta f} \right>_{\mathbf{z}} \cdot \mathbf{z}^* -  \left< \frac{\delta \mathbf{y}}{\delta f} \right>_{\mathbf{y}} \cdot \mathbf{y}^* \right).$$

The mean values $\left< s \right>_{QP} = \left< \psi_\Sigma \right>_{QP} = \left< \psi_{mi} \right>_{QP} = \left< \psi_\Sigma \right>_{QPs} = \left< \psi_{mi} \right>_{QPs} = 0$, since in both models the mean value of $p_i$ is 0  for all particles and the positions of small particles in the $QP$ model have as the mean value the position of the large particle $Q$. Note that $s = \sum_i (q_i - Q)/N$, $\psi_\Sigma=\sum_i p_i/(N \sum_j m_j)$, and $\psi_{mi} = \sum_i p_i/(N m_i)$. This leaves only the calculation of $\left< \psi_{mi} \right>_{\Sigma}, \left< \psi_\Sigma \right>_{mi}$, which after a few tedious operations leads to the mean values
$$\left< \psi_{mi} \right>_{\Sigma}\psi_{mi}^* = \left< \psi_\Sigma \right>_{mi} \psi_\Sigma^* = - \frac{T}{N \sum_j m_j} \psi_\Sigma^* \psi_{mi}^*.$$

\section{On the equivalence of spectral and Fourier definitions of $(I-\Delta)^{1/2}$}\label{appen:spectral}

We briefly justify that the definition of $(I-\Delta)^{1/2}$ via the spectral calculus associated with the self-adjoint operator $-\Delta$ agrees with its realization as a Fourier multiplier on $\mathbb{R}^n$. This follows from two complementary observations: the spectral mapping theorem and the fact that Fourier modes diagonalize the Laplacian.

We begin with the simpler setting, where $-\Delta$ has purely discrete spectrum (this occurs, for example, for sufficiently regular and bounded domains with suitable Robin boundary conditions). Denoting by $\{(\lambda_k,\varphi_k)\}_{k=1}^\infty$ the eigenpairs, the spectral theorem yields the resolution
\begin{equation*}
-\Delta = \sum_{k=1}^\infty \lambda_k P_k, \qquad P_k f = \langle f,\varphi_k\rangle \varphi_k.
\end{equation*}
For any Borel function $g$, the (Borel) functional calculus defines
\begin{equation*}
g(-\Delta) = \sum_{k=1}^\infty g(\lambda_k) P_k,
\end{equation*}
by the virtue of resolution of the identity. This is consistent with the spectral mapping theorem, which asserts that the spectrum of $g(-\Delta)$ is precisely $\{g(\lambda_k)\}$. In particular, for $g(\lambda) = (1+\lambda)^{1/2}$, one obtains
\begin{equation*}
(I-\Delta)^{1/2} f = \sum_{k=1}^\infty (1+\lambda_k)^{1/2} \langle f,\varphi_k\rangle \varphi_k,
\end{equation*}
i.e. the operator acts diagonally by multiplying each spectral component by $(1+\lambda_k)^{1/2}$.

On $\mathbb{R}^n$, the spectrum of $-\Delta$ is continuous and the role of the eigenbasis is played by the family of plane waves $e^{2\pi i \rr\cdot \xxi}$, which satisfy
\begin{equation*}
-\Delta e^{2\pi i \rr\cdot \xxi} = 4\pi^2 |\xxi|^2 e^{2\pi i \rr\cdot \xxi}.
\end{equation*}
The Fourier transform thus provides the spectral representation of $-\Delta$, with spectral variable $\lambda = 4\pi^2 |\xxi|^2$. In this representation, the functional calculus implies that $g(-\Delta)$ acts by multiplication with $g(4\pi^2 |\xxi|^2)$. Consequently, for $g(\lambda) = (1+\lambda)^{1/2}$, one finds
\begin{equation*}
\widehat{(I-\Delta)^{1/2} f}(\xxi) = \sqrt{1+4\pi^2 |\xxi|^2}\,\hat{f}(\xxi),
\end{equation*}
which is precisely the Fourier multiplier definition from the main text.

\paragraph{}
In summary, both constructions are manifestations of the same principle: once $-\Delta$ is diagonalized (either discretely via eigenfunctions or continuously via Fourier modes), the operator $g(-\Delta)$ is obtained by applying $g$ to the corresponding spectral values. This establishes the equivalence of the spectral and Fourier definitions of $(I-\Delta)^{1/2}$. 

% \subsection{Additional figures}

% \begin{figure}[ht] 
% \centering
% \includegraphics[scale=0.3]{images/Plot_Sols+Diss.png}
% \includegraphics[scale=0.3]{images/Plot_Sols+S_e.png}
% \caption{Contour plot (julia) of the function $S_e (T, u_T)$ (right) and the "distribution" $exp{(-S_e (T, u_T))}$ (left) together with the solutions of \eqref{eqn:DSdT=0} ($u_T = \frac{1}{cosh{(T)}}$, white) and \eqref{eqn:dudTNabla} (yellow). Initial $u_0 = 1$}
% \end{figure}

% \begin{figure}[ht] 
% \centering
% \includegraphics[scale=0.3]{images/Plot_Se_trajs2.png}
% \caption{Plot of the function $S_e (T, u_T)$ together with the solutions of \eqref{eqn:DSdT=0} ($u_T = \frac{1}{cosh{(T)}}$, green), \eqref{eqn:dudTNabla} (black), and \eqref{eqn:dudTNabla_HJ} (red, $C_2=0$). Initial $u_0 = 1$.}
% \end{figure}    

%\bibliographystyle{apalike}
%\bibliography{library}

\end{document}